\documentclass[aps,
prapp,
reprint,
noeprint,
superscriptaddress,
amsmath,
amssymb,
]{revtex4-2}
\usepackage[english]{babel}
\usepackage{amsmath}
\usepackage{graphicx}
\usepackage[colorlinks=true, allcolors=blue]{hyperref}
\usepackage{times}
\usepackage{multirow}
\usepackage{bm}
\usepackage{braket}
\usepackage{graphicx}
\usepackage{booktabs}
\usepackage{adjustbox} 
\usepackage{listings}%
\usepackage{CJKutf8}

\newcommand{\NUAAPhy}{\affiliation{1}{College of Physics, Nanjing University of Aeronautics and Astronautics, Nanjing 211106, China.}}
\newcommand{\NUAACS}{\affiliation{1}{College of Computer Science and Technology, Nanjing University of Aeronautics and Astronautics, Nanjing 211106, China.}}
\newcommand{\CM}{\affiliation{1}{China Mobile (Suzhou) Software Technology Co., Ltd., Suzhou 215163, China}}
\newcommand{\SIOM}{\affiliation{1}{Shanghai Institute of Optics and Fine Mechanics, Chinese Academy of Sciences, Shanghai 201800, China}}

\begin{document}

\title{Learning quantum disentanglement scheduling from reduced states via modular hybrid policies}

\author{Y.-X. Xiao}
\thanks{These authors contributed equally to this work.}
\affiliation{\NUAAPhy}
\affiliation{\NUAACS}

\author{J.-Z. Han}
\thanks{These authors contributed equally to this work.}
\affiliation{\CM}
\affiliation{\NUAAPhy}

\author{Z. Zheng}
\affiliation{\CM}

\author{Z.-H. Zhang}
\affiliation{\CM}

\author{M. Xue}
\affiliation{\NUAAPhy}
\affiliation{Key Laboratory of Aerospace Information Materials and Physics (NUAA), MIIT, Nanjing 211106, China}

\author{J. Li}
\affiliation{\NUAAPhy}
\affiliation{Key Laboratory of Aerospace Information Materials and Physics (NUAA), MIIT, Nanjing 211106, China}

\author{X. Lv}
\affiliation{\SIOM}

\date{\today}

\begin{abstract}
Quantum control with restricted state access is central to near-term quantum devices, where full wave-function information is unavailable. We study this problem through multiqubit disentanglement scheduling from partial observations, where a controller receives only two-qubit reduced density matrices and selects which qubit pair to disentangle at each step. We introduce a modular hybrid quantum--classical policy framework consisting of classical preprocessing, a parameterized quantum circuit as a compact nonlinear latent block, and classical
postprocessing for pair-selection probabilities. Benchmarking 4-, 5-, and 6-qubit tasks, we find that preprocessing is the dominant factor governing performance under reduced-state observations, while the quantum module provides a conditional compact representation whose utility depends on the input features and model budget. We further identify a performance--efficiency trade-off across policy families and find that increasing circuit width is generally more useful
than increasing depth. These results provide practical design principles for hybrid policies in reduced-information quantum control.
\end{abstract}

\maketitle

\section{Introduction}
Quantum disentanglement is a fundamental control problem in multiqubit systems and is closely connected to state initialization, circuit synthesis, and efficient state preparation in quantum information processing~\cite{nielsen2010quantum,1629135,PhysRevA.83.032302,PhysRevLett.129.230504,PhysRevLett.119.170503,doi:10.1126/science.aax9743,Madjarov_2020,doi:10.1126/science.adf4272,Iqbal_2024,Bluvstein_2023}. More broadly, disentangling operations are also relevant to circuit simplification and robust operation on near-term quantum devices~\cite{RevModPhys.95.045005,PhysRevA.63.062309,PhysRevA.91.062306,PhysRevA.91.062307}. Existing studies have mainly investigated entanglement manipulation~\cite{dehaghani2025optimizingmaximallyentangledstate} through matrix-product-state descriptions~\cite{PhysRevB.98.235163,PRXQuantum.3.030343}, and using optimal-control constructions~\cite{Lu_2024}, while a general and scalable strategy for disentangling unknown many-body states remains much less developed. In practice, this difficulty is rooted in the fact that effective disentangling decisions depend on global correlation structure, whereas full-state descriptions scale exponentially with system size and are generally inaccessible in realistic settings~\cite{Cramer_2010}. This limitation has recently motivated learning-based approaches, including reinforcement-learning-based disentanglement from partial observations, which offer a new route to sequential control in multiqubit systems~\cite{tashev2024reinforcementlearningdisentanglemultiqubit}.

Reinforcement learning has been applied broadly in quantum science, including quantum state control~\cite{PhysRevX.8.031084,PhysRevX.8.031086,PhysRevB.98.224305,Dalgaard_2020,Haug_2021,Mackeprang_2020,yao2020policygradientbasedquantum,yao2020noiserobustendtoendquantumcontrol,PhysRevLett.126.060401,niu2018universalquantumcontroldeep} and circuit compilation tasks~\cite{PhysRevLett.125.170501,PhysRevLett.127.110502,Moro_2021,He_2021,fösel2021quantumcircuitoptimizationdeep}. At the same time, quantum computing has also been increasingly integrated into machine-learning and reinforcement-learning frameworks, giving rise to a growing body of work on quantum reinforcement learning and hybrid quantum--classical learning architectures~\cite{meyer2024surveyquantumreinforcementlearning,4579244,alexeev2025artificial,NEURIPS2021_eec96a7f,Skolik_2022}. In these approaches, parameterized quantum circuit (PQC) are typically used as trainable components within a larger learning pipeline, where they can provide compact nonlinear transformations and potentially enrich the representation capability of the overall model~\cite{jin2025ppoqproximalpolicyoptimization,Wu_2025,NEURIPS2021_eec96a7f}. Such developments suggest that hybrid quantum--classical architectures may offer a useful new design space for reinforcement-learning policies, especially in structured control problems.

Once the problem is cast in this form, the central question is no longer whether reinforcement learning can solve the task, but which policy architectures are most effective under reduced-state observations. The difficulty is structural: the policy must infer discrete control actions from incomplete local information about multiqubit correlations, making representation design central to successful learning. This makes representation design the main bottleneck of reduced-state policy learning. Existing Transformer-based policies already demonstrate that this task can be solved~\cite{tashev2024reinforcementlearningdisentanglemultiqubit}, but they also point to a practical architectural issue: as the task grows harder, strong encoders tend to require substantial model capacity, raising questions of parameter efficiency, extensibility, and architectural bias~\cite{li2020trainlargecompressrethinking}. These considerations motivate a more systematic study of preprocessing design, compact hybrid latent modules, and the resulting performance--efficiency trade-off. 

In this work, we investigate partially observed quantum disentanglement from this architecture-oriented perspective. We introduce a unified modular hybrid policy framework consisting of preprocessing, parameterized quantum-circuit, and postprocessing modules, enabling controlled comparisons under a common reinforcement-learning pipeline. Within this framework, the preprocessing stage extracts scheduling-relevant information from local reduced states, the PQC acts as a compact nonlinear latent transformation, and the postprocessing stage maps the resulting representation to pair-selection decisions. Using 4-, 5-, and 6-qubit disentanglement tasks, we show that preprocessing design is the dominant factor governing performance under reduced-state observations, whereas the quantum module plays a more conditional role as a compact intermediate representation block. We further identify a clear performance--efficiency trade-off across model families and find that, within the hybrid module, increasing PQC width is generally more effective than increasing depth under the settings considered here. These results provide an architecture-oriented view of partially observed quantum disentanglement and offer practical guidance for modular policy design in reduced-information quantum control.

\begin{figure*}[t]
    \centering
     \includegraphics[width=0.96\textwidth]{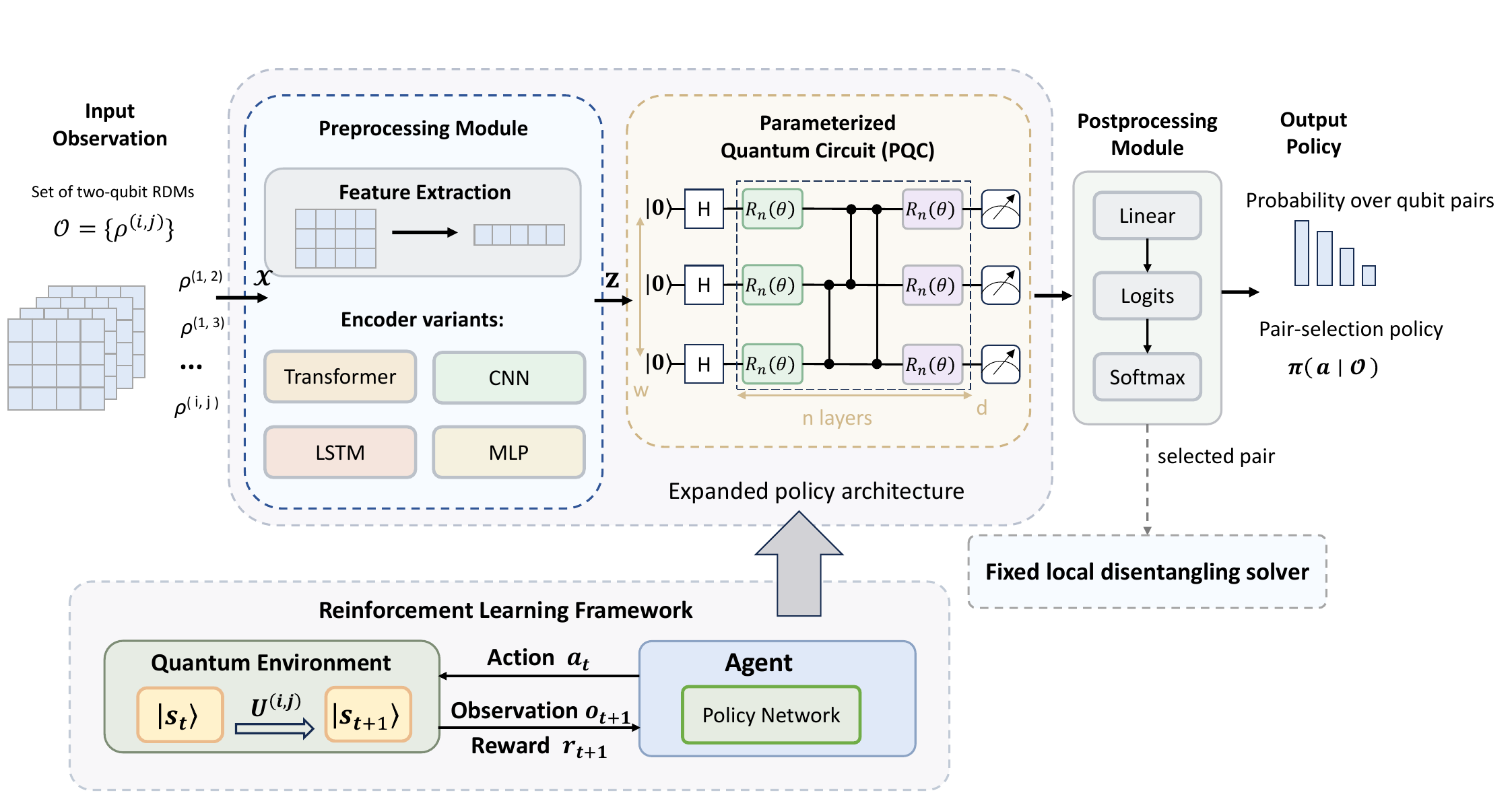}
    \caption{\textbf{Overall architecture of the proposed hybrid quantum--classical policy network within the partially observed quantum disentanglement framework.}
    The lower panel shows the reinforcement-learning interaction loop between the quantum environment and the agent, where the policy receives reduced-state observations and outputs an action at each step. The upper panel expands the policy network into three modules: a preprocessing module, a PQC module, and a postprocessing module. The preprocessing stage extracts pairwise features from the set of two-qubit reduced density matrices and maps them to a compact latent representation. The PQC acts as a compact nonlinear transformation block, and the postprocessing module converts the resulting features into action logits and a probability distribution over candidate qubit pairs. The selected pair is then passed to the fixed local disentangling solver.}
    \label{fig:hybrid_policy_framework}
\end{figure*}                           

\section{Modular hybrid quantum--classical policy framework}
\subsection{Overall architecture of the hybrid policy network}

Under the partially observed disentanglement setting introduced above, the policy network takes as input the collection of two-qubit reduced density matrices (RDMs) and outputs a probability distribution over candidate qubit pairs. We adopt a hybrid quantum--classical policy architecture composed of three successive modules: a preprocessing module, a PQC module, and a postprocessing module. The overall data flow can be written as
\begin{equation}
\mathcal{O}
\;\xrightarrow{\;f_{\mathrm{pre}}\;}
\mathbf{z}
\;\xrightarrow{\;f_{\mathrm{PQC}}\;}
\mathbf{m}
\;\xrightarrow{\;f_{\mathrm{post}}\;}
\pi(\cdot \mid \mathcal{O}),
\end{equation}
where $\mathcal{O}$ denotes the set of local reduced-state observations, $\mathbf{z}$ is the latent representation produced by the preprocessing module, $\mathbf{m}$ is the measurement vector returned by the PQC, and $\pi(\cdot \mid \mathcal{O})$ is the policy over all qubit pairs.

Figure~\ref{fig:hybrid_policy_framework} summarizes the proposed policy pipeline. The preprocessing module converts the collection of two-qubit reduced density matrices into a compact latent representation, the PQC module applies a trainable nonlinear transformation in this latent space~\cite{10.5555/3505464.3505499,9144562}, and the postprocessing module maps the resulting features to a probability distribution over candidate qubit pairs. The selected pair is then passed to a fixed local disentangling solver.

This decomposition makes the role of each component explicit. The preprocessing module is responsible for extracting useful global scheduling information from partial local observations, the PQC acts as a compact intermediate representation block, and the postprocessing module converts the transformed features into action preferences. Because the local solver is fixed throughout ~\cite{PhysRevA.59.3320}, the framework allows us to isolate architectural effects from gate-construction details.

\subsection{Preprocessing module for reduced-state observations}

The input observation consists of $L(L-1)/2$ two-qubit reduced density matrices. To make them compatible with neural-network processing, each reduced density matrix $\rho^{(i,j)}$ is first converted into a real-valued feature vector,
\begin{equation}
\mathbf{x}_{ij} = g\!\left(\rho^{(i,j)}\right),
\end{equation}
where $g(\cdot)$ denotes the feature extraction and vectorization procedure. The full observation is then represented as a collection of pairwise features,
\begin{equation}
\mathcal{X}=\{\mathbf{x}_{ij}\mid 1\le i<j\le L\}.
\end{equation}

The preprocessing module maps this collection to a compact latent representation,
\begin{equation}
\mathbf{z}=f_{\mathrm{pre}}(\mathcal{X}),
\end{equation}
which is then passed to the PQC module. Its role is to extract informative local features, model dependencies among different qubit pairs, and compress the input into a low-dimensional representation suitable for subsequent quantum processing.

Different preprocessing models induce different representation biases. 
In this work, we instantiate $f_{\mathrm{pre}}$ with multiple candidate architectures and compare their performance under the same downstream hybrid policy framework. This allows us to study how the choice of preprocessing model affects disentanglement success, gate efficiency, and parameter budget in a controlled setting.

\subsection{Parameterized quantum circuit module}

The PQC module~\cite{jin2025ppoqproximalpolicyoptimization,NEURIPS2021_eec96a7f} serves as a compact nonlinear transformation block inside the policy network. It operates on the latent representation produced by the preprocessing module rather than on the physical quantum state being disentangled. In other words, the PQC is used here as an internal representation module of the policy, not as a model of the underlying system dynamics.

Let $n_q$ denote the number of qubits in the PQC circuit. The latent vector $\mathbf{z}$ is first projected to a PQC-compatible input and encoded into a quantum state. Starting from the initial state $\ket{0}^{\otimes n_q}$, we apply a hardware-efficient circuit composed of repeated parameterized blocks. In our implementation, each layer contains single-qubit rotations and \emph{entangling gates},
\begin{equation}
R_y(\theta^{(y)}_{l,q})R_z(\theta^{(z)}_{l,q}),
\end{equation}
for qubit $q$ in layer $l$, followed by CZ gates between neighboring qubits. The circuit depth is controlled by the number of repeated layers.

After the final layer, each qubit is measured in the Pauli-$Z$ basis, yielding the measurement vector
\begin{equation}
\mathbf{m}
=
\left(
\langle Z_1\rangle,\langle Z_2\rangle,\ldots,\langle Z_{n_q}\rangle
\right).
\end{equation}
This vector provides a compact quantum-transformed representation of the latent features. In the experiments, we further study how different PQC configurations, including the number of qubits and the circuit depth, affect the performance of the overall policy.

\subsection{Postprocessing module and model variants}

The postprocessing module maps the PQC measurement vector to logits over the action space and outputs the final policy distribution. Concretely, we use a lightweight classical network
\begin{equation}
\mathbf{h}=f_{\mathrm{post}}(\mathbf{m}),
\end{equation}
followed by softmax normalization,
\begin{equation}
\pi(a\mid\mathcal{O})
=
\frac{\exp(h_a)}
{\sum_{a'\in A}\exp(h_{a'})},
\qquad a\in A,
\end{equation}
where $A$ is the set of candidate qubit pairs.

Within this unified framework, we compare multiple model variants by changing the preprocessing module and the PQC configuration while keeping the remaining components fixed. All variants share the same RL environment, reward function, local disentangling solver, and termination criterion. Therefore, differences in performance can be attributed directly to architectural choices rather than to changes in the control setting. This design enables a controlled study of how preprocessing models and PQC configurations influence the performance--efficiency trade-off in partially observed quantum disentanglement.

\section{Problem Setup}

\subsection{Partially observed quantum disentanglement task}

We study the problem of disentangling an unknown pure state of an $L$-qubit system through a sequence of two-qubit operations. Given an entangled input state, the goal is to reach a fully separable product state using as few disentangling steps as possible.

The task is formulated as a sequential decision-making problem with a clear separation between global scheduling and local gate synthesis~\cite{tashev2024reinforcementlearningdisentanglemultiqubit,KAELBLING199899}. At each step, the agent selects a qubit pair, while the corresponding two-qubit disentangling operation is generated by a fixed local solver. Hence, the policy does not predict continuous gate parameters; it only determines which pair should be processed next.

Under this formulation, the controller operates solely on partial observations derived from local reduced states. The objective of learning is therefore to infer an effective disentangling schedule from incomplete local information.

\subsection{Observation space based on two-qubit RDMs}

Directly using the full $L$-qubit state as the policy input is impractical because its representation size grows exponentially with the number of qubits~\cite{Cramer_2010,DONG2022243}. Moreover, full-state information is typically unavailable in realistic settings~\cite{Sotnikov_2022,alexeev2025artificial}. To obtain a compact and physically meaningful observation interface, we restrict the policy input to the set of two-qubit reduced density matrices.

For each unordered qubit pair $(i,j)$ with $1 \le i < j \le L$, let $\rho^{(i,j)}$ denote the corresponding two-qubit reduced density matrix obtained by tracing out all remaining qubits. The observation provided to the agent is then the collection
\begin{equation}
\mathcal{O} = \left\{ \rho^{(i,j)} \;\middle|\; 1 \le i < j \le L \right\}.
\end{equation}
This observation space scales quadratically with the system size and provides local correlation information that is directly relevant to pairwise disentangling operations.

In practice, each reduced density matrix is converted into a real-valued feature representation before being fed into the policy network. The resulting set of pairwise features forms the input to the preprocessing module introduced in the next section. This reduced-observation formulation preserves the essential local structure of the problem while avoiding the prohibitive complexity of full-state descriptions.

\subsection{Action space and local disentangling solver}

At each decision step, the agent selects one qubit pair from the candidate set
\begin{equation}
\mathcal{A} = \left\{ (i,j) \mid 1 \le i < j \le L \right\},
\end{equation}
whose cardinality is
\begin{equation}
|\mathcal{A}| = \frac{L(L-1)}{2}.
\end{equation}
The policy therefore defines a probability distribution over all possible qubit pairs,
\begin{equation}
\pi(\cdot \mid \mathcal{O}) : \mathcal{O} \rightarrow [0,1]^{|\mathcal{A}|},
\qquad
\sum_{a \in \mathcal{A}} \pi(a \mid \mathcal{O}) = 1.
\end{equation}

Once a pair $(i,j)$ is selected, the corresponding two-qubit disentangling operation is not predicted by the network. Instead, it is generated analytically by a fixed local solver based on the reduced density matrix of that pair. In this way, the learning problem remains purely discrete: the policy learns \emph{which pair to act on}, while the local solver determines \emph{how to act on it}~\cite{tashev2024reinforcementlearningdisentanglemultiqubit}. This design keeps the control pipeline consistent across all compared models and allows us to isolate the effect of policy architecture from that of gate construction.

Because all methods in this work share the same action space and the same local solver, differences in performance can be attributed directly to the quality of the learned scheduling policy. This property is important for our later comparisons among different preprocessing encoders and prediction-head designs.

\subsection{Reward design and evaluation objectives}

The reward is designed to encourage rapid entanglement reduction while discouraging unnecessarily long disentangling sequences. Following the disentanglement setting considered in this work, we define the stepwise reward as
\begin{equation}
R(s_t,a_t,s_{t+1})
\!=\!
\sum_{j=1}^{L}\!
\frac{
S_{\mathrm{ent}}[\rho_t^{(j)}]
\!-\!
S_{\mathrm{ent}}[\rho_{t+1}^{(j)}]
}{
\max\!\left(
S_{\mathrm{ent}}[\rho_t^{(j)}],
S_{\mathrm{ent}}[\rho_{t+1}^{(j)}]
\right)
}
\!-\!
n(s_{t+1}),
\label{eq:reward}
\end{equation}
where $\rho_t^{(j)}$ is the one-qubit reduced density matrix of qubit $j$ at step $t$, $S_{\mathrm{ent}}[\rho]$ denotes the single-qubit entanglement entropy, and $n(s_{t+1})$ is the number of qubits that remain entangled after the transition. The first term rewards normalized entropy reduction, while the second term penalizes states that are still far from complete disentanglement.

In simulations, we evaluate the learned policies using four metrics: the disentanglement success rate, the average gate count over successful episodes, the standard deviation of gate count, and the number of trainable parameters. These metrics jointly reflect policy effectiveness, circuit efficiency, stability, and model complexity.

\begin{figure*}
    \centering
    \includegraphics[width=0.90\textwidth]{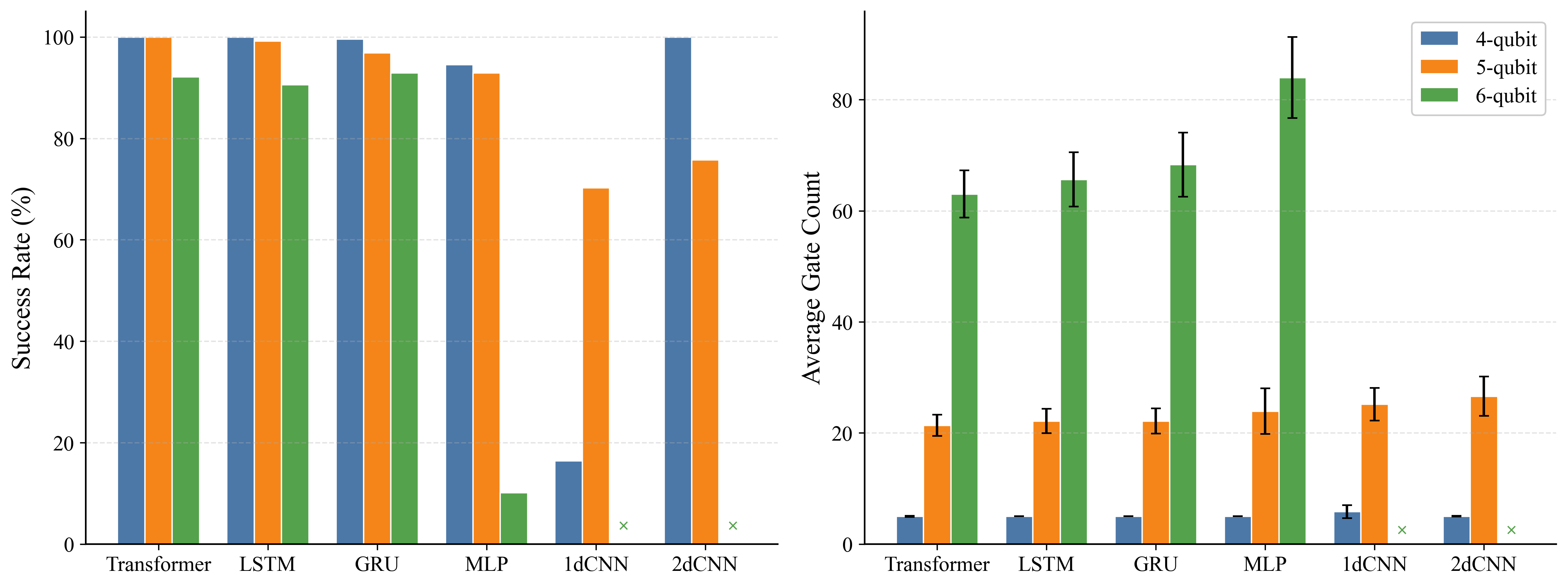}
    \caption{\textbf{Comparison of disentanglement performance across preprocessing models under fully entangled 4-, 5-, and 6-qubit settings.}
    The statistics are computed over 500 randomly generated fully entangled initial states for each system size. The left panel shows the disentanglement success rate within a maximum budget of 128 steps. The right panel reports the average disentanglement step count over successful episodes, with error bars indicating the standard deviation. The results show that all models perform strongly in the 4-qubit setting, whereas the performance gap becomes increasingly pronounced as the system size grows, with the clearest separation appearing for 6 qubits. Stronger preprocessing architectures achieve not only higher success rates but also lower circuit costs, indicating that improved reduced-state representations benefit both effectiveness and efficiency. Markers near the horizontal axis indicate cases where no successful disentanglement episode was obtained.}
\label{fig:fully_entangled_summary}
    \label{fig:fully_entangled_summary}
\end{figure*}

\section{Numerical Experiments}

\subsection{Setup}
We evaluate the proposed framework on partially observed disentanglement tasks with $L\in\{4,5,6\}$. To isolate architectural effects, we compare multiple preprocessing models under the same reinforcement-learning pipeline: all variants share the same reward function, local disentangling solver, termination criterion, critic network, and environment-side components. During evaluation, we further apply a simple action-selection refinement to suppress degenerate repeated decisions; details are deferred to the appendix.

We report four metrics: disentanglement success rate, average gate count over successful episodes, gate-count standard deviation, and the number of trainable parameters. These respectively quantify policy effectiveness, circuit cost, stability, and model complexity.

Our numerical study addresses three questions: how preprocessing models scale from 4 to 6 qubits, where the main performance differences emerge in the 6-qubit setting, and whether these differences are explained primarily by parameter count or by genuine architectural trade-offs.

\subsection{Comparison across qubit numbers under fully entangled settings}

We first benchmark the models on fully entangled 4-, 5- and 6-qubit tasks, which provide the most demanding instances in our evaluation protocol. Figure~\ref{fig:fully_entangled_summary} reports the disentanglement success rate and average disentanglement step count for different preprocessing architectures. A clear overall trend is that all models perform strongly in the 4-qubit setting, whereas the separation between architectures becomes increasingly pronounced as the system size grows, with the largest gap appearing for 6 qubits.

The simultaneous separation in success rate and average gate count is informative. It shows that stronger preprocessing models do not merely make successful disentanglement more likely, but also lead to more efficient disentangling schedules once success is achieved. In this sense, the observed gain is not only one of reliability but also one of control quality: better preprocessing enables the policy to identify more effective action sequences and thus reduce the circuit cost of successful episodes.

This scaling behavior supports the architecture-oriented interpretation of the task. Under reduced-state observations, the policy must construct globally useful decisions from incomplete local entanglement information. For small systems, this inference problem remains relatively mild, so even weaker architectures can often produce workable policies. As the number of qubits increases, however, the disentangling process becomes more globally constrained, and the representation bottleneck becomes progressively more severe. Figure~\ref{fig:fully_entangled_summary} therefore indicates that the main limiting factor at larger scales is not generic task difficulty alone, but the ability of the policy architecture to organize local reduced-state information into a globally coherent disentangling strategy.

\begin{figure*}
    \centering
    \includegraphics[width=0.85\textwidth]{fig3_patternwise_comparison_6q.png}
    \caption{\textbf{Pattern-wise comparison on 6-qubit disentanglement tasks.}
    Different preprocessing models are compared across representative 6-qubit initial-state patterns. The statistics and plotting conventions are the same as in Fig.~\ref{fig:fully_entangled_summary}. In the pattern labels, consecutive \(R\)'s without a hyphen denote qubits belonging to the same Haar-random entangled block, while hyphens separate different blocks. Model differences remain small on several structured patterns but become much more pronounced on the hardest patterns, indicating that the performance gap observed at larger scales is concentrated on high-complexity disentanglement instances rather than distributed uniformly across all tasks.}
\label{fig:patternwise_6q}
    \label{fig:patternwise_6q}
\end{figure*}

\subsection{Fine-grained analysis on 6-qubit entanglement patterns}

To identify where the main architectural differences arise, we next resolve the 6-qubit results by entanglement pattern. Figure~\ref{fig:patternwise_6q} reports the average gate count and success rate across representative 6-qubit initial-state families. A clear feature of the data is that the performance gap is strongly pattern dependent rather than uniform: on several structured patterns, most models achieve near-perfect success rates with comparable gate counts, whereas much larger differences appear on the hardest patterns.

The pattern labels represent different global organizations of entanglement, ranging from more decomposable block structures to more strongly coupled instances. When the entanglement structure is relatively regular, the local observation interface already contains sufficiently clear cues for effective scheduling, so even weaker architectures can often produce workable policies. By contrast, in the hardest patterns the local information becomes less directly actionable, and the policy must integrate multiple weak cues into a globally coherent disentangling strategy. It is precisely in this regime that stronger preprocessing models retain higher success rates and lower circuit costs.

The degradation observed at larger scales is not a uniform loss of performance over all inputs, but is concentrated on high-complexity disentanglement instances. This shows that the main role of the preprocessing architecture is not simply to improve average behavior in an undifferentiated sense, but to sustain global decision quality when the entanglement structure places the greatest demand on reduced-state representation learning.

\begin{figure}[t]
    \centering
    \includegraphics[width=1\columnwidth]{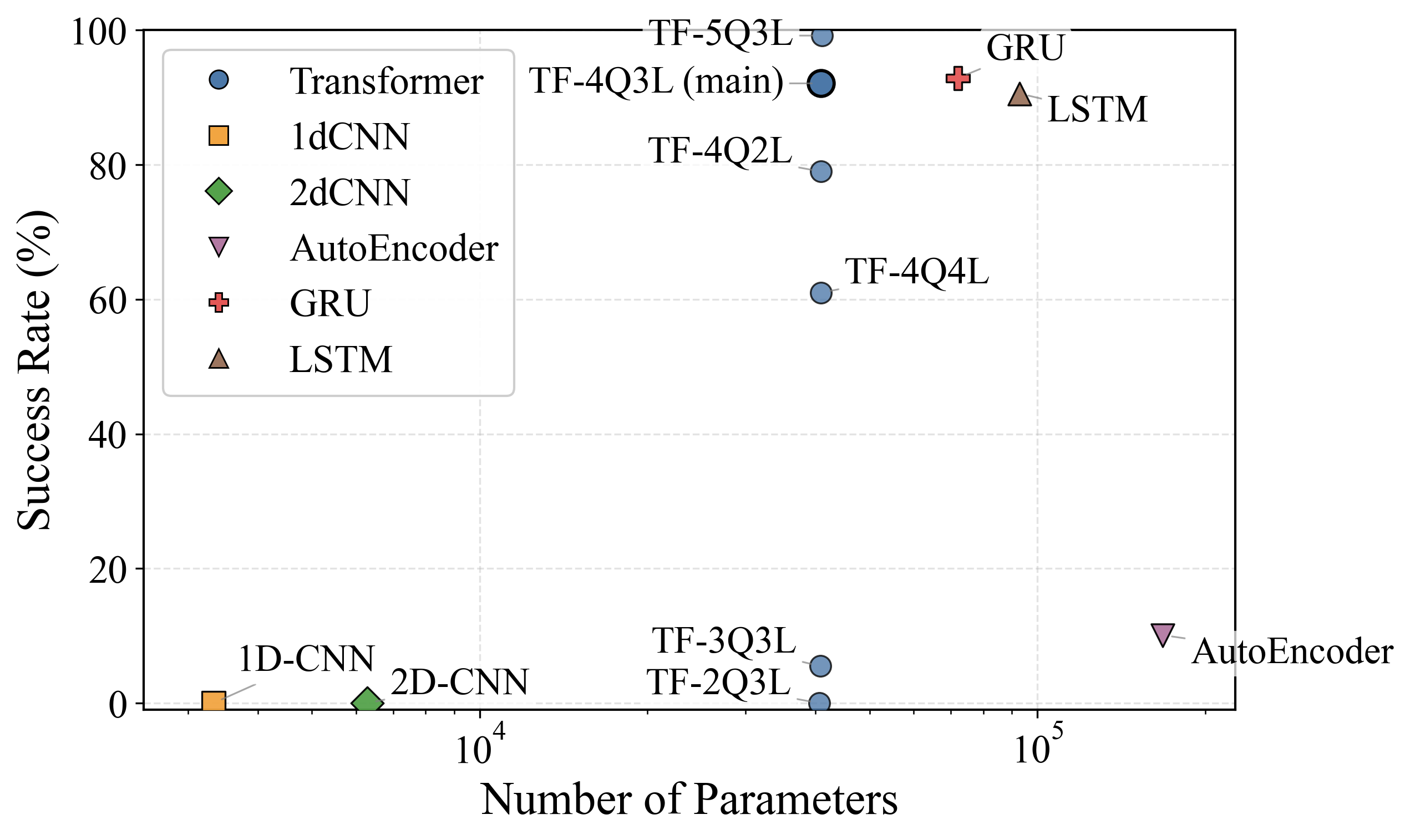}
    \caption{\textbf{Performance--parameter trade-off across preprocessing models and PQC configurations.}
    Different preprocessing models and representative PQC configurations are compared in the success-rate--parameter plane. The horizontal axis shows the total number of trainable parameters of the complete model, while the vertical axis reports the disentanglement success rate on the 6-qubit fully entangled task. The statistics are computed under the same evaluation setting as in Fig.~\ref{fig:fully_entangled_summary}. For the comparison across preprocessing models, all variants use the same PQC backbone with 4 qubits and 3 layers. In addition, the points labeled TF-\(n\)Q\(m\)L correspond to Transformer-based models in which the PQC width and depth are varied, including different PQC qubit numbers under a fixed 3-layer setting and different PQC depths under a fixed 4-qubit setting. The results show that model performance is not determined solely by parameter count: architectures with similar parameter budgets can still exhibit markedly different success rates. At the same time, varying the PQC configuration within a fixed preprocessing backbone can shift the model substantially in the trade-off plane, indicating that both preprocessing design and PQC structure contribute to the final performance.}
\label{fig:architecture_tradeoff}
    \label{fig:architecture_tradeoff}
\end{figure}

\subsection{Architecture trade-offs across preprocessing models and PQC configurations}

To further clarify how architectural choices govern disentanglement performance, we examine the trade-off between success rate and model complexity across both preprocessing models and PQC configurations. Figure~\ref{fig:architecture_tradeoff} places different model variants in the same performance--parameter plane, thereby providing a unified view of architecture-dependent trade-offs rather than treating preprocessing design and PQC design as disconnected ablations. Two types of variation are shown simultaneously: the comparison across different preprocessing architectures under a common PQC backbone, and the comparison across different PQC widths and depths within the Transformer-based backbone.

The first observation is that the performance ranking cannot be explained simply by model size. Although larger models may in some cases achieve stronger results, the gain is neither monotonic nor proportional to the parameter budget. Architectures with comparable numbers of trainable parameters can still exhibit markedly different success rates, while increases in parameter count do not always translate into proportional performance improvements. This indicates that raw model capacity alone is insufficient to account for the observed differences. Instead, the way in which the preprocessing architecture organizes, compresses, and transforms reduced-state information appears to play the central role.

The second observation is that a similar trade-off structure persists even within a fixed preprocessing backbone when the PQC configuration is varied. Table~\ref{tab:pqc_ablation} makes this point more precise for the 6-qubit fully entangled setting. A purely classical MLP head already provides a reasonable baseline, but a suitably configured hybrid head can achieve both higher success rate and lower average gate count. At the same time, very small PQCs perform poorly or even fail completely, showing that the introduction of a quantum module by itself does not guarantee useful internal representations. The PQC therefore should not be viewed as a universally superior replacement for a classical head, but rather as a conditional intermediate module whose usefulness depends on whether sufficient latent representational capacity is available.

A more specific conclusion concerns the relative roles of PQC width and depth. Under a fixed Transformer preprocessing module, increasing the number of PQC qubits leads to a much more substantial improvement than simply stacking additional circuit layers. In particular, widening the PQC from 4 to 5 qubits significantly improves success rate, whereas increasing the depth beyond the best 4-qubit configuration does not yield further gains and may even degrade performance. This behavior suggests that, in the present task, representational width is a more relevant resource than circuit depth. Taken together, Fig.~\ref{fig:architecture_tradeoff} and Table~\ref{tab:pqc_ablation} support the interpretation that policy performance is controlled jointly by preprocessing quality and PQC structure, with preprocessing remaining the dominant factor and the PQC acting as a compact nonlinear latent transformation block.

\begin{table}[t]
\centering
\caption{\textbf{Effect of prediction heads and PQC configurations under the 6-qubit fully entangled setting.}
The preprocessing encoder is fixed. Success rate is computed as the fraction of successfully disentangled episodes. Avg.\ Gates is reported over successful episodes only. For configurations with no successful episode, the average gate count is undefined and is marked as ``--''.}
\label{tab:pqc_ablation}
\resizebox{0.48\textwidth}{!}{
\begin{tabular}{lccccc}
\toprule
Prediction Head & PQC Qubits & PQC Depth & Success (\%) & Avg.\ Gates & Std. \\
\midrule
MLP head        & -- & -- & 78.91 & 63.85 & 4.43 \\
Hybrid head     & 2  & 3  & 0.00  & --    & --   \\
Hybrid head     & 3  & 3  & 0.78  & 80.00 & 1.00 \\
Hybrid head     & 4  & 2  & 78.91 & 65.08 & 5.51 \\
Hybrid head     & 4  & 3  & 92.19 & 63.02 & 4.27 \\
Hybrid head     & 4  & 4  & 60.94 & 68.33 & 5.74 \\
Hybrid head     & 5  & 3  & 99.22 & 60.74 & 3.83 \\
\bottomrule
\end{tabular}
}
\end{table}

\subsection{Training dynamics from single-qubit entropy evolution}

Beyond the final success rates and disentanglement costs, the training process itself provides additional insight into how different architectures form disentangling policies. Figure~\ref{fig:entropy_dynamics} reports the evolution of the single-qubit average entanglement-entropy distribution at representative epochs during training. A common trend across all models is that successful learning is accompanied by a progressive shift of the distribution toward lower entropy values, reflecting the gradual formation of policies that more consistently reduce local entanglement.

At the same time, the dynamics differ markedly across preprocessing architectures. Stronger models drive the entropy distribution toward the low-entropy regime more rapidly and more stably, whereas weaker models exhibit slower shifts, broader distributions, or persistent high-entropy tails even at later stages of training. These differences are significant because they show that the architectural gap is not only a matter of final converged performance, but is already visible in how efficiently the policy representation develops during learning.

This dynamical behavior is consistent with the architecture-oriented interpretation advanced throughout this work. Under reduced-state observations, the policy must construct globally useful scheduling decisions from incomplete local information. When the preprocessing module is effective, this internal representation emerges earlier and supports more reliable entropy reduction throughout training. Conversely, weaker preprocessing modules struggle to organize the same local information into a sufficiently coherent decision structure, leading to slower or less stable progress toward disentanglement. Figure~\ref{fig:entropy_dynamics} therefore provides complementary evidence that preprocessing quality acts as the main bottleneck in reduced-state policy learning.

\begin{figure*}[t]
    \centering
    \includegraphics[width=0.92\textwidth]{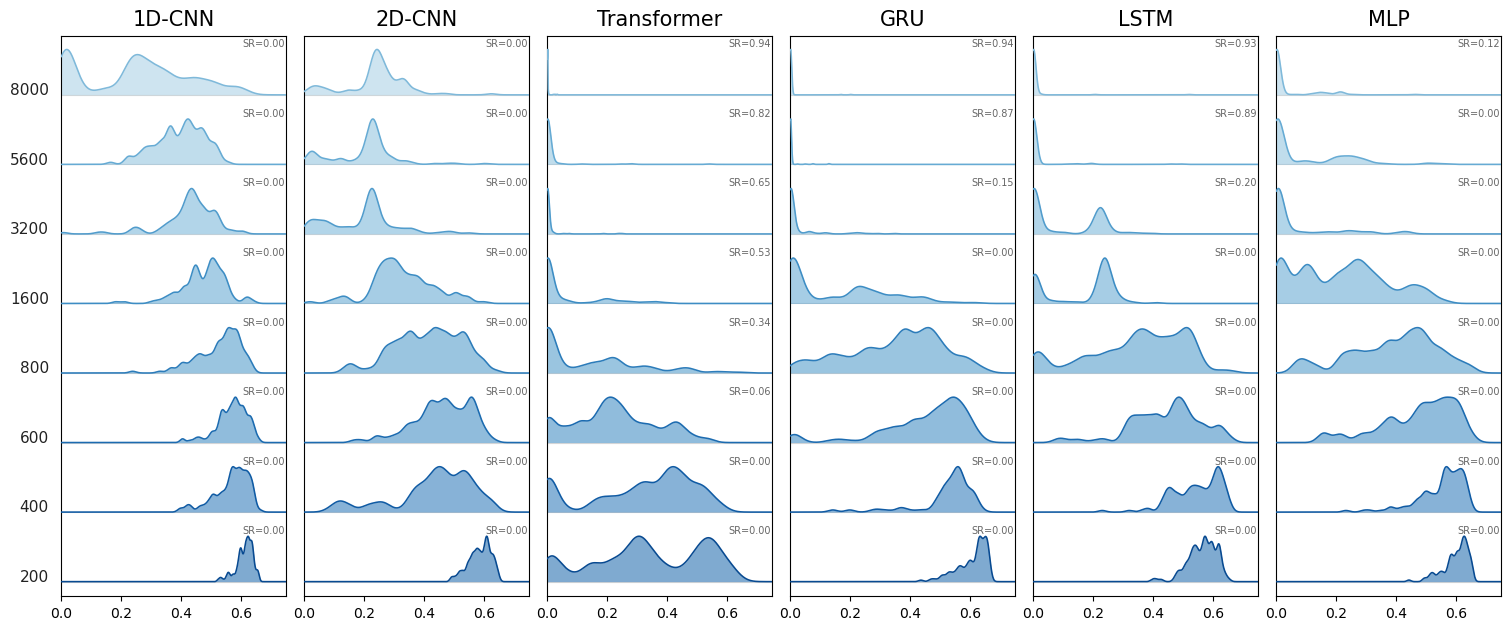}
    \caption{\textbf{Training dynamics from single-qubit entropy evolution.}
    Single-qubit average entanglement-entropy distributions are shown at representative training epochs over a total of 8000 iterations. The evaluation uses the same set of 500 initial states for all models, under the same testing setting as in the preceding experiments. Each column corresponds to one preprocessing architecture, and each row corresponds to one selected training epoch. The value ``SR'' shown in the upper-right corner of each panel denotes the solve rate at that epoch, where an episode is counted as successfully disentangled if the average single-qubit entropy is below $\epsilon=10^{-3}$. The progressive shift of the distribution toward lower entropy values reflects the formation of more effective disentangling policies. Stronger preprocessing models tend to exhibit faster, sharper, and more stable concentration into the low-entropy regime, whereas weaker models remain broader or retain substantial high-entropy mass for longer training periods.}
\label{fig:entropy_dynamics}
    \label{fig:entropy_dynamics}
\end{figure*}

\section{Conclusion and Discussion}

We have studied partially observed quantum disentanglement from an architecture-oriented perspective within a unified hybrid quantum--classical policy framework. By decomposing the policy into preprocessing, parameterized quantum-circuit (PQC), and postprocessing modules, we performed controlled comparisons under a common reinforcement-learning pipeline and isolated the architectural factors that govern policy performance. Across 4-, 5-, and 6-qubit tasks, the results consistently show that preprocessing design is the dominant determinant of performance under reduced-state observations, and that its importance becomes increasingly pronounced as the task grows more difficult.

These findings clarify the nature of the control problem considered here. When the agent has access only to collections of two-qubit reduced density matrices, successful disentanglement depends on whether incomplete local information can be transformed into globally useful scheduling decisions. From this perspective, the dominant role of preprocessing is not incidental, but reflects a genuine representation bottleneck in partially observed quantum control. The strongest architectural differences emerge precisely in those regimes where local reduced-state cues must be integrated into a globally coherent policy.

The present study also clarifies the role of the hybrid quantum module. The PQC is best understood not as a universally superior substitute for a classical prediction head, but as a compact nonlinear latent transformation block whose usefulness depends on the surrounding architecture. In particular, the observation that increasing PQC width is generally more beneficial than increasing depth suggests that latent representational capacity is more valuable than additional circuit layering in the present setting. More broadly, the results reveal a clear performance--efficiency trade-off across model families: better performance is not determined by parameter count alone, but by how effectively the architecture organizes, compresses, and transforms reduced-state information before action selection.

Several extensions deserve further study. It will be important to test whether the same architectural trends persist for larger systems, broader state families, and more realistic settings with noise, imperfect reduced-state estimation, and hardware constraints. It would also be interesting to move beyond fixed local solvers and investigate joint optimization of scheduling and gate synthesis within a unified framework. More generally, the present results suggest that hybrid quantum--classical control policies should be designed as modular systems in which representation learning, compact nonlinear transformation, and action decoding are optimized together rather than in isolation.

\appendix

\section{Interpretation of the Two-Qubit Local Update Rule}
\label{app:local_optimal_gate_interpretation}

In this appendix, we give an intuitive interpretation of the two-qubit local update rule used in the disentangling procedure and explain why the construction can be carried out entirely from the two-qubit reduced density matrix of the selected pair.

A point that should be emphasized at the outset is that the unitary $U^{(i,j)}$ is ultimately applied to the full $L$-qubit pure state $|\psi\rangle$. The use of the two-qubit reduced density matrix
\begin{equation}
\rho^{(i,j)}=\mathrm{tr}_{\mathrm{rest}}\!\left(|\psi\rangle\langle\psi|\right)
\end{equation}
is a way of formulating the local update rule for the chosen pair $(i,j)$ in a reduced description. Since the gate acts nontrivially only on qubits $i$ and $j$, the information needed to construct this local step is contained in $\rho^{(i,j)}$.

To motivate the update, one may adopt the following local objective: for the selected pair $(i,j)$, choose a two-qubit unitary that minimizes the entanglement internal to the two-qubit mixed state $\rho^{(i,j)}$. For a general two-qubit mixed state, the entanglement of formation is
\begin{equation}
E_f\!\left[\rho^{(i,j)}\right] = h(x),
\end{equation}
where
\begin{equation}
h(x)=-x\log x-(1-x)\log(1-x),
\end{equation}
and
\begin{equation}
x=\frac{1+\sqrt{1-C^2\!\left(\rho^{(i,j)}\right)}}{2}.
\end{equation}
Here $C(\rho^{(i,j)})$ is the concurrence. In the standard Wootters form,
\begin{equation}
C(\rho)=\max\!\left\{0,\sqrt{\nu_1}-\sqrt{\nu_2}-\sqrt{\nu_3}-\sqrt{\nu_4}\right\},
\end{equation}
where $\nu_k$ are the eigenvalues, in decreasing order, of
\begin{equation}
\rho\,\tilde{\rho},
\qquad
\tilde{\rho}=(\sigma_y\otimes\sigma_y)\rho^*(\sigma_y\otimes\sigma_y).
\end{equation}
Since $E_f$ is a monotonically increasing function of $C$, minimizing the concurrence is equivalent to minimizing the entanglement of formation.

Now let the spectral decomposition of the selected reduced state be
\begin{equation}
\rho^{(i,j)}=\sum_{k=1}^{4}\lambda_k|\phi_k\rangle\langle\phi_k|.
\end{equation}
Choose a two-qubit unitary $U^{(i,j)}$ such that
\begin{equation}
U^{(i,j)}|\phi_k\rangle=|b_k\rangle,
\end{equation}
where $\{|b_k\rangle\}$ is the computational basis
\begin{equation}
\{|00\rangle,|01\rangle,|10\rangle,|11\rangle\},
\end{equation}
up to ordering. Then the transformed reduced state becomes
\begin{equation}
\tilde{\rho}^{(i,j)}
=
U^{(i,j)}\rho^{(i,j)}U^{(i,j)\dagger}
=
\mathrm{diag}(\lambda_1,\lambda_2,\lambda_3,\lambda_4)
\end{equation}
in the computational basis.

At this point, the key observation is simple: a density matrix that is diagonal in the computational basis is a classical probabilistic mixture of the product states
\begin{equation}
|00\rangle,\ |01\rangle,\ |10\rangle,\ |11\rangle.
\end{equation}
Therefore,
\begin{equation}
\tilde{\rho}^{(i,j)}
=
\lambda_1|00\rangle\langle 00|
+\lambda_2|01\rangle\langle 01|
+\lambda_3|10\rangle\langle 10|
+\lambda_4|11\rangle\langle 11|
\end{equation}
is separable as a two-qubit state. Hence its concurrence vanishes,
\begin{equation}
C\!\left(\tilde{\rho}^{(i,j)}\right)=0,
\end{equation}
and consequently
\begin{equation}
E_f\!\left[\tilde{\rho}^{(i,j)}\right]=0.
\end{equation}
In this precise sense, the diagonalizing update completely removes the entanglement internal to the selected two-qubit reduced state.

This also yields a simple interpretation for the one-qubit marginals. Because $\tilde{\rho}^{(i,j)}$ is diagonal in the computational basis, the corresponding single-qubit reduced density matrices $\tilde{\rho}^{(i)}$ and $\tilde{\rho}^{(j)}$ are diagonal as well. Thus the local update suppresses both two-qubit coherence and single-qubit coherence in the chosen basis, making the selected subsystem locally as classical as possible within this update rule.

Several remarks are in order. First, the gate is explicitly state dependent: since $U^{(i,j)}$ is constructed from the instantaneous reduced density matrix $\rho^{(i,j)}$, the optimal local update for the same pair generally changes during the evolution. Second, the update is not unique: different orderings of the eigenvectors, or further composition with local basis conventions, may lead to equivalent diagonal forms of the reduced state. Finally, local optimality does not imply global optimality. Although this construction eliminates the internal entanglement of the selected two-qubit reduced state in one step, a sequence of such local updates need not produce the shortest or globally optimal disentangling circuit for the full $L$-qubit system. Its role is instead to provide a well-defined and computationally tractable local rule within the sequential disentangling protocol.

\section{Representative Disentangling Trajectories}
\label{app:representative_trajectories}

\begin{figure*}[htbp]
    \centering
    \includegraphics[width=\textwidth]{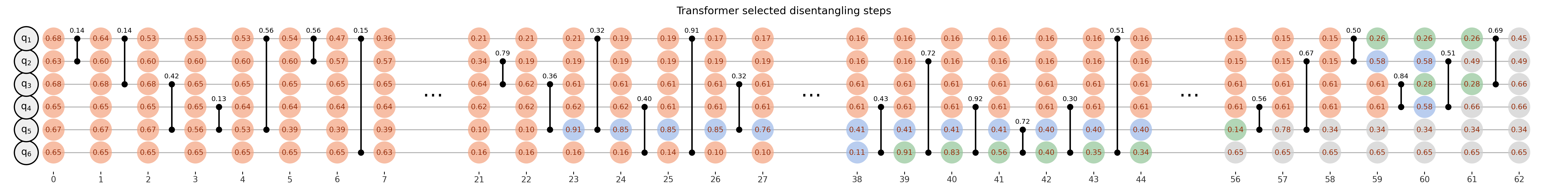}
    \caption{Disentangling steps for the Transformer-based policy on a six-qubit test instance.}
    \label{fig:app_transformer_steps}
\end{figure*}

\begin{figure*}[htbp]
    \centering
    \includegraphics[width=\textwidth]{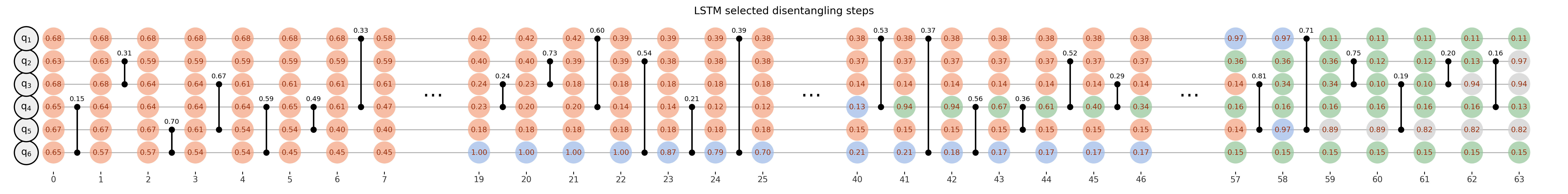}
    \caption{Disentangling steps for the LSTM-based policy on a six-qubit test instance.}
    \label{fig:app_lstm_steps}
\end{figure*}

\begin{figure*}[htbp]
    \centering
    \includegraphics[width=\textwidth]{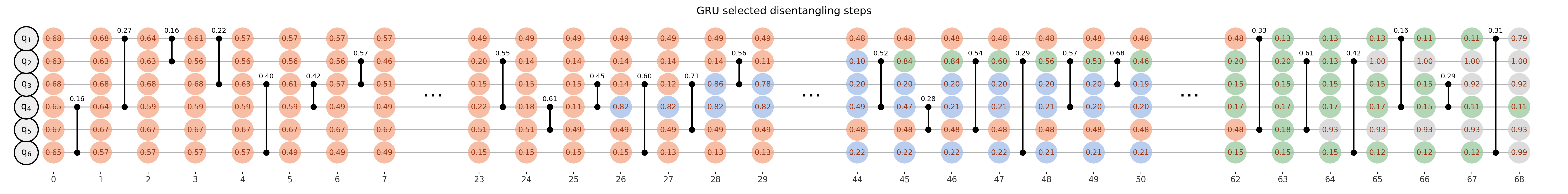}
    \caption{Disentangling steps for the GRU-based policy on a six-qubit test instance.}
    \label{fig:app_gru_steps}
\end{figure*}

\begin{figure*}[htbp]
    \centering
    \includegraphics[width=\textwidth]{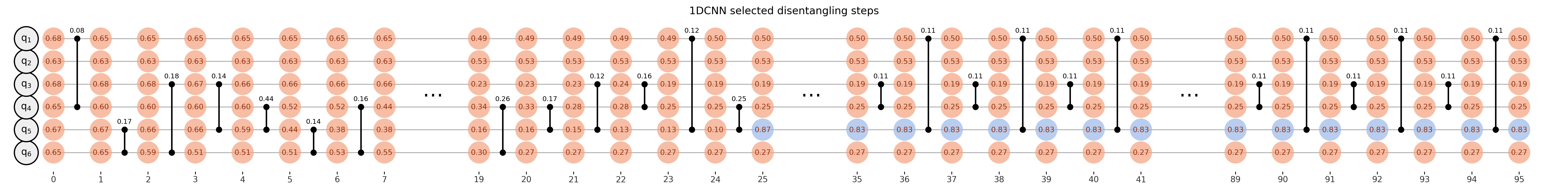}
    \caption{Disentangling steps for the 1D-CNN-based policy on a six-qubit test instance.}
    \label{fig:app_cnn1d_steps}
\end{figure*}

\begin{figure*}[htbp]
    \centering
    \includegraphics[width=\textwidth]{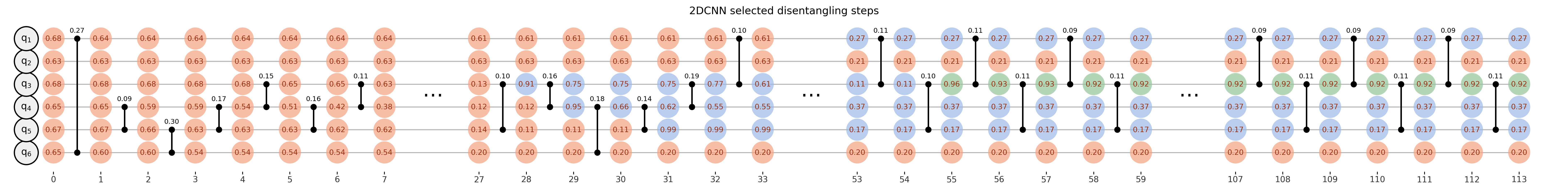}
    \caption{Disentangling steps for the 2D-CNN-based policy on a six-qubit test instance.}
    \label{fig:app_cnn2d_steps}
\end{figure*}

\begin{figure*}[htbp]
    \centering
    \includegraphics[width=\textwidth]{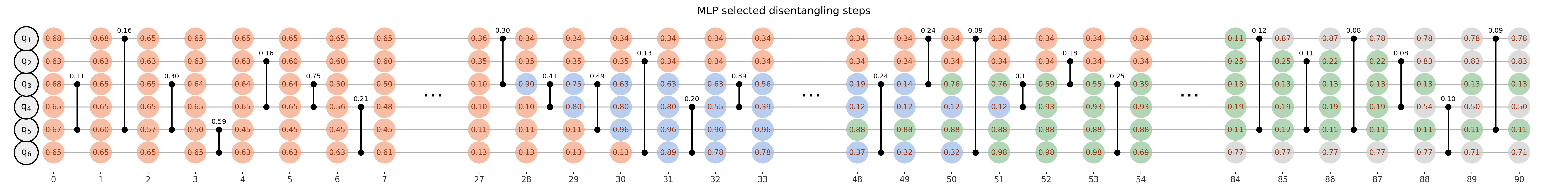}
    \caption{Disentangling steps for the MLP-based policy on a six-qubit test instance.}
    \label{fig:app_mlp_steps}
\end{figure*}

To further illustrate how different preprocessing architectures affect the sequential disentangling behavior, we provide representative disentangling trajectories for six-qubit test instances in Fig.~\ref{fig:app_transformer_steps}--Fig.~\ref{fig:app_mlp_steps}. Since the full trajectories are often long and visually dense, we retain only several representative step ranges for each model. These selected segments are intended to highlight the characteristic evolution patterns of different policies rather than to reproduce the full episode in exhaustive detail.

In all figures, each column represents the system state after one disentangling action, and the black vertical connector indicates the selected two-qubit pair acted upon at that step. The colored circles report the single-qubit entropies for each qubit, with the color scale indicating the corresponding order of magnitude. This visualization makes it possible to track how local disentangling actions reorganize the entanglement distribution across qubits over time.

Overall, the selected trajectories are consistent with the quantitative results reported in the main text. Stronger preprocessing models typically exhibit a more structured progression: after an initial redistribution stage, they gradually suppress the residual entanglement of a smaller subset of qubits and eventually approach a nearly disentangled state. By contrast, weaker models often reach an intermediate configuration in which part of the entanglement has been reduced, but the remaining residual structure is not further resolved, leading to stagnation or failure within the allowed horizon.

Figure~\ref{fig:app_transformer_steps} shows a representative successful trajectory produced by the Transformer-based policy. The selected step ranges suggest a multi-stage disentangling process. In the early stage, the policy does not immediately drive all qubits toward low entropy, but instead performs a sequence of local updates that gradually reshapes the entanglement distribution. This behavior indicates that the policy first reorganizes the partially observed reduced-state information into a more favorable intermediate structure.

A clearer reduction of residual entanglement appears in the middle stage, where several qubits begin to transition from order $10^{0}$ entropy to lower orders. The late stage is characterized by a more targeted cleanup process: only a small subset of qubits still carries visible residual entropy, and the remaining actions progressively suppress these components. This trajectory is consistent with the view that the Transformer encoder is effective at extracting global scheduling information from local reduced-density-matrix observations.

A representative LSTM trajectory is shown in Fig.~\ref{fig:app_lstm_steps}. Compared with the Transformer case, the LSTM policy also exhibits a staged reduction pattern, but its progress appears more local and sequential. In the earlier segment, the policy begins by weakening entanglement on part of the system while leaving a few qubits at relatively high entropy. This suggests that the policy is able to exploit temporal regularities in the action sequence, but does so in a more incremental fashion.

In the middle segment, one observes a clearer separation between qubits that have already been substantially disentangled and those that still dominate the residual entropy. The later segment shows that the policy continues to reduce these remaining components and eventually reaches a successful near-disentangled configuration. Thus, the LSTM policy still supports successful long-horizon control, although its disentangling pattern appears somewhat less globally coordinated than that of the Transformer.

Figure~\ref{fig:app_gru_steps} presents a representative trajectory for the GRU-based policy. The overall pattern indicates a relatively mild reduction in the early stage, followed by a more substantial decrease at later steps. In other words, the GRU policy appears to require a longer preparatory phase before entering a regime in which the residual entanglement is more effectively compressed.

The selected middle and late segments show that once a favorable intermediate structure is formed, the policy can indeed drive several qubits into lower entropy regimes. However, compared with the Transformer and LSTM cases, the transition is somewhat less smooth, and the final stage still exhibits a certain degree of irregularity in the residual entropy profile. Even so, the trajectory remains qualitatively consistent with successful disentanglement, and it supports the interpretation that the GRU can realize delayed but ultimately effective disentangling progress.

Figure~\ref{fig:app_cnn1d_steps} shows a representative trajectory for the 1D-CNN policy. In the early stage, the policy is able to reduce part of the entanglement and move the system away from its initial high-entropy configuration. However, this initial progress is followed by a pronounced stagnation regime. The selected later segments show that the entropy profile becomes nearly stationary, with a persistent set of qubits retaining non-negligible residual entropy.

This pattern suggests that the 1D-CNN encoder can capture certain local structural cues from the observation, but lacks the representational capacity needed to support long-range coordination over the full action sequence. As a result, the policy reaches an intermediate partially disentangled configuration but fails to further resolve the remaining entanglement within the available episode length. This stagnation behavior is also consistent with the unsuccessful outcomes observed for this model in the main experiments.

The representative 2D-CNN trajectory in Fig.~\ref{fig:app_cnn2d_steps} exhibits a pattern similar in spirit to that of the 1D-CNN, although the intermediate reduction stage is somewhat more pronounced. In the first retained segment, the policy already begins to redistribute the entanglement across qubits. In the subsequent segment, several qubits enter lower-entropy regimes, indicating that the model has captured some useful structural information from the partially observed input.

Nevertheless, the late-stage segment reveals that the policy settles into another persistent plateau rather than continuing toward full disentanglement. A stable residual entropy pattern remains, and additional actions do not substantially improve it. This suggests that, although the 2D-CNN encoder provides somewhat richer local feature extraction than the 1D-CNN, it still does not produce representations sufficient for robust long-horizon disentangling decisions in the present setting.

Figure~\ref{fig:app_mlp_steps} presents a representative successful trajectory for the MLP-based policy. Relative to the CNN-based models, the MLP policy continues to make progress after the intermediate stage rather than remaining trapped in a plateau. In the earlier retained segment, the policy starts to suppress the entropies of a subset of qubits, while in the middle segments it further compresses the remaining entanglement and drives more qubits into the lower-order regime.

The later segment indicates that the MLP policy can eventually complete the disentangling process, although the reduction pattern is less smooth and less globally organized than in the Transformer case. In this sense, the MLP occupies an intermediate position among the tested models: it is capable of successful disentanglement, but its sequential control behavior appears less structured than that produced by the stronger sequence-oriented encoders.

Taken together, these representative trajectories provide a more direct view of the qualitative differences among preprocessing architectures. Transformer-, LSTM-, and GRU-based policies all display sustained sequential progress, although with different degrees of smoothness and coordination. Among them, the Transformer trajectory appears the most globally organized, while recurrent models exhibit a somewhat more incremental pattern. In contrast, the CNN-based policies tend to reduce entanglement only up to an intermediate level and then remain trapped in a stable residual configuration. The MLP, while simpler, still retains sufficient flexibility to complete the disentangling process in representative successful cases.

These observations support the main claim of this work: under partial observations constructed from two-qubit reduced density matrices, preprocessing is not merely a front-end implementation choice, but a central component that shapes how the policy organizes local information into effective long-horizon disentangling actions.

\bibliography{paper} 

\begin{thebibliography}{49}%
\makeatletter
\providecommand \@ifxundefined [1]{%
 \@ifx{#1\undefined}
}%
\providecommand \@ifnum [1]{%
 \ifnum #1\expandafter \@firstoftwo
 \else \expandafter \@secondoftwo
 \fi
}%
\providecommand \@ifx [1]{%
 \ifx #1\expandafter \@firstoftwo
 \else \expandafter \@secondoftwo
 \fi
}%
\providecommand \natexlab [1]{#1}%
\providecommand \enquote  [1]{``#1''}%
\providecommand \bibnamefont  [1]{#1}%
\providecommand \bibfnamefont [1]{#1}%
\providecommand \citenamefont [1]{#1}%
\providecommand \href@noop [0]{\@secondoftwo}%
\providecommand \href [0]{\begingroup \@sanitize@url \@href}%
\providecommand \@href[1]{\@@startlink{#1}\@@href}%
\providecommand \@@href[1]{\endgroup#1\@@endlink}%
\providecommand \@sanitize@url [0]{\catcode `\\12\catcode `\$12\catcode
  `\&12\catcode `\#12\catcode `\^12\catcode `\_12\catcode `\%12\relax}%
\providecommand \@@startlink[1]{}%
\providecommand \@@endlink[0]{}%
\providecommand \url  [0]{\begingroup\@sanitize@url \@url }%
\providecommand \@url [1]{\endgroup\@href {#1}{\urlprefix }}%
\providecommand \urlprefix  [0]{URL }%
\providecommand \Eprint [0]{\href }%
\providecommand \doibase [0]{https://doi.org/}%
\providecommand \selectlanguage [0]{\@gobble}%
\providecommand \bibinfo  [0]{\@secondoftwo}%
\providecommand \bibfield  [0]{\@secondoftwo}%
\providecommand \translation [1]{[#1]}%
\providecommand \BibitemOpen [0]{}%
\providecommand \bibitemStop [0]{}%
\providecommand \bibitemNoStop [0]{.\EOS\space}%
\providecommand \EOS [0]{\spacefactor3000\relax}%
\providecommand \BibitemShut  [1]{\csname bibitem#1\endcsname}%
\let\auto@bib@innerbib\@empty
\bibitem [{\citenamefont {Nielsen}\ and\ \citenamefont
  {Chuang}(2010)}]{nielsen2010quantum}%
  \BibitemOpen
  \bibfield  {author} {\bibinfo {author} {\bibfnamefont {M.~A.}\ \bibnamefont
  {Nielsen}}\ and\ \bibinfo {author} {\bibfnamefont {I.~L.}\ \bibnamefont
  {Chuang}},\ }\href
  {https://www.cambridge.org/highereducation/books/quantum-computation-and-quantum-information/01E10196D0A682A6AEFFEA52D53BE9AE#overview}
  {\emph {\bibinfo {title} {Quantum computation and quantum information}}}\
  (\bibinfo  {publisher} {Cambridge university press},\ \bibinfo {year}
  {2010})\BibitemShut {NoStop}%
\bibitem [{\citenamefont {Shende}\ \emph {et~al.}(2006)\citenamefont {Shende},
  \citenamefont {Bullock},\ and\ \citenamefont {Markov}}]{1629135}%
  \BibitemOpen
  \bibfield  {author} {\bibinfo {author} {\bibfnamefont {V.}~\bibnamefont
  {Shende}}, \bibinfo {author} {\bibfnamefont {S.}~\bibnamefont {Bullock}},\
  and\ \bibinfo {author} {\bibfnamefont {I.}~\bibnamefont {Markov}},\
  }\bibfield  {title} {\bibinfo {title} {Synthesis of quantum-logic circuits},\
  }\href {https://doi.org/10.1109/TCAD.2005.855930} {\bibfield  {journal}
  {\bibinfo  {journal} {IEEE Transactions on Computer-Aided Design of
  Integrated Circuits and Systems}\ }\textbf {\bibinfo {volume} {25}},\
  \bibinfo {pages} {1000} (\bibinfo {year} {2006})}\BibitemShut {NoStop}%
\bibitem [{\citenamefont {Plesch}\ and\ \citenamefont
  {Brukner}(2011)}]{PhysRevA.83.032302}%
  \BibitemOpen
  \bibfield  {author} {\bibinfo {author} {\bibfnamefont {M.}~\bibnamefont
  {Plesch}}\ and\ \bibinfo {author} {\bibfnamefont {i.~c.~v.}\ \bibnamefont
  {Brukner}},\ }\bibfield  {title} {\bibinfo {title} {Quantum-state preparation
  with universal gate decompositions},\ }\href
  {https://doi.org/10.1103/PhysRevA.83.032302} {\bibfield  {journal} {\bibinfo
  {journal} {Phys. Rev. A}\ }\textbf {\bibinfo {volume} {83}},\ \bibinfo
  {pages} {032302} (\bibinfo {year} {2011})}\BibitemShut {NoStop}%
\bibitem [{\citenamefont {Zhang}\ \emph {et~al.}(2022)\citenamefont {Zhang},
  \citenamefont {Li},\ and\ \citenamefont {Yuan}}]{PhysRevLett.129.230504}%
  \BibitemOpen
  \bibfield  {author} {\bibinfo {author} {\bibfnamefont {X.-M.}\ \bibnamefont
  {Zhang}}, \bibinfo {author} {\bibfnamefont {T.}~\bibnamefont {Li}},\ and\
  \bibinfo {author} {\bibfnamefont {X.}~\bibnamefont {Yuan}},\ }\bibfield
  {title} {\bibinfo {title} {Quantum state preparation with optimal circuit
  depth: Implementations and applications},\ }\href
  {https://doi.org/10.1103/PhysRevLett.129.230504} {\bibfield  {journal}
  {\bibinfo  {journal} {Phys. Rev. Lett.}\ }\textbf {\bibinfo {volume} {129}},\
  \bibinfo {pages} {230504} (\bibinfo {year} {2022})}\BibitemShut {NoStop}%
\bibitem [{\citenamefont {Eldredge}\ \emph {et~al.}(2017)\citenamefont
  {Eldredge}, \citenamefont {Gong}, \citenamefont {Young}, \citenamefont
  {Moosavian}, \citenamefont {Foss-Feig},\ and\ \citenamefont
  {Gorshkov}}]{PhysRevLett.119.170503}%
  \BibitemOpen
  \bibfield  {author} {\bibinfo {author} {\bibfnamefont {Z.}~\bibnamefont
  {Eldredge}}, \bibinfo {author} {\bibfnamefont {Z.-X.}\ \bibnamefont {Gong}},
  \bibinfo {author} {\bibfnamefont {J.~T.}\ \bibnamefont {Young}}, \bibinfo
  {author} {\bibfnamefont {A.~H.}\ \bibnamefont {Moosavian}}, \bibinfo {author}
  {\bibfnamefont {M.}~\bibnamefont {Foss-Feig}},\ and\ \bibinfo {author}
  {\bibfnamefont {A.~V.}\ \bibnamefont {Gorshkov}},\ }\bibfield  {title}
  {\bibinfo {title} {Fast quantum state transfer and entanglement
  renormalization using long-range interactions},\ }\href
  {https://doi.org/10.1103/PhysRevLett.119.170503} {\bibfield  {journal}
  {\bibinfo  {journal} {Phys. Rev. Lett.}\ }\textbf {\bibinfo {volume} {119}},\
  \bibinfo {pages} {170503} (\bibinfo {year} {2017})}\BibitemShut {NoStop}%
\bibitem [{\citenamefont {Omran}\ \emph {et~al.}(2019)\citenamefont {Omran},
  \citenamefont {Levine}, \citenamefont {Keesling}, \citenamefont {Semeghini},
  \citenamefont {Wang}, \citenamefont {Ebadi}, \citenamefont {Bernien},
  \citenamefont {Zibrov}, \citenamefont {Pichler}, \citenamefont {Choi},
  \citenamefont {Cui}, \citenamefont {Rossignolo}, \citenamefont {Rembold},
  \citenamefont {Montangero}, \citenamefont {Calarco}, \citenamefont {Endres},
  \citenamefont {Greiner}, \citenamefont {Vuletić},\ and\ \citenamefont
  {Lukin}}]{doi:10.1126/science.aax9743}%
  \BibitemOpen
  \bibfield  {author} {\bibinfo {author} {\bibfnamefont {A.}~\bibnamefont
  {Omran}}, \bibinfo {author} {\bibfnamefont {H.}~\bibnamefont {Levine}},
  \bibinfo {author} {\bibfnamefont {A.}~\bibnamefont {Keesling}}, \bibinfo
  {author} {\bibfnamefont {G.}~\bibnamefont {Semeghini}}, \bibinfo {author}
  {\bibfnamefont {T.~T.}\ \bibnamefont {Wang}}, \bibinfo {author}
  {\bibfnamefont {S.}~\bibnamefont {Ebadi}}, \bibinfo {author} {\bibfnamefont
  {H.}~\bibnamefont {Bernien}}, \bibinfo {author} {\bibfnamefont {A.~S.}\
  \bibnamefont {Zibrov}}, \bibinfo {author} {\bibfnamefont {H.}~\bibnamefont
  {Pichler}}, \bibinfo {author} {\bibfnamefont {S.}~\bibnamefont {Choi}},
  \bibinfo {author} {\bibfnamefont {J.}~\bibnamefont {Cui}}, \bibinfo {author}
  {\bibfnamefont {M.}~\bibnamefont {Rossignolo}}, \bibinfo {author}
  {\bibfnamefont {P.}~\bibnamefont {Rembold}}, \bibinfo {author} {\bibfnamefont
  {S.}~\bibnamefont {Montangero}}, \bibinfo {author} {\bibfnamefont
  {T.}~\bibnamefont {Calarco}}, \bibinfo {author} {\bibfnamefont
  {M.}~\bibnamefont {Endres}}, \bibinfo {author} {\bibfnamefont
  {M.}~\bibnamefont {Greiner}}, \bibinfo {author} {\bibfnamefont
  {V.}~\bibnamefont {Vuletić}},\ and\ \bibinfo {author} {\bibfnamefont
  {M.~D.}\ \bibnamefont {Lukin}},\ }\bibfield  {title} {\bibinfo {title}
  {Generation and manipulation of schrödinger cat states in rydberg atom
  arrays},\ }\href {https://doi.org/10.1126/science.aax9743} {\bibfield
  {journal} {\bibinfo  {journal} {Science}\ }\textbf {\bibinfo {volume}
  {365}},\ \bibinfo {pages} {570} (\bibinfo {year} {2019})}\BibitemShut
  {NoStop}%
\bibitem [{\citenamefont {Madjarov}\ \emph {et~al.}(2020)\citenamefont
  {Madjarov}, \citenamefont {Covey}, \citenamefont {Shaw}, \citenamefont
  {Choi}, \citenamefont {Kale}, \citenamefont {Cooper}, \citenamefont
  {Pichler}, \citenamefont {Schkolnik}, \citenamefont {Williams},\ and\
  \citenamefont {Endres}}]{Madjarov_2020}%
  \BibitemOpen
  \bibfield  {author} {\bibinfo {author} {\bibfnamefont {I.~S.}\ \bibnamefont
  {Madjarov}}, \bibinfo {author} {\bibfnamefont {J.~P.}\ \bibnamefont {Covey}},
  \bibinfo {author} {\bibfnamefont {A.~L.}\ \bibnamefont {Shaw}}, \bibinfo
  {author} {\bibfnamefont {J.}~\bibnamefont {Choi}}, \bibinfo {author}
  {\bibfnamefont {A.}~\bibnamefont {Kale}}, \bibinfo {author} {\bibfnamefont
  {A.}~\bibnamefont {Cooper}}, \bibinfo {author} {\bibfnamefont
  {H.}~\bibnamefont {Pichler}}, \bibinfo {author} {\bibfnamefont
  {V.}~\bibnamefont {Schkolnik}}, \bibinfo {author} {\bibfnamefont {J.~R.}\
  \bibnamefont {Williams}},\ and\ \bibinfo {author} {\bibfnamefont
  {M.}~\bibnamefont {Endres}},\ }\bibfield  {title} {\bibinfo {title}
  {High-fidelity entanglement and detection of alkaline-earth rydberg atoms},\
  }\href {https://doi.org/10.1038/s41567-020-0903-z} {\bibfield  {journal}
  {\bibinfo  {journal} {Nature Physics}\ }\textbf {\bibinfo {volume} {16}},\
  \bibinfo {pages} {857–861} (\bibinfo {year} {2020})}\BibitemShut {NoStop}%
\bibitem [{\citenamefont {Holland}\ \emph {et~al.}(2023)\citenamefont
  {Holland}, \citenamefont {Lu},\ and\ \citenamefont
  {Cheuk}}]{doi:10.1126/science.adf4272}%
  \BibitemOpen
  \bibfield  {author} {\bibinfo {author} {\bibfnamefont {C.~M.}\ \bibnamefont
  {Holland}}, \bibinfo {author} {\bibfnamefont {Y.}~\bibnamefont {Lu}},\ and\
  \bibinfo {author} {\bibfnamefont {L.~W.}\ \bibnamefont {Cheuk}},\ }\bibfield
  {title} {\bibinfo {title} {On-demand entanglement of molecules in a
  reconfigurable optical tweezer array},\ }\href
  {https://doi.org/10.1126/science.adf4272} {\bibfield  {journal} {\bibinfo
  {journal} {Science}\ }\textbf {\bibinfo {volume} {382}},\ \bibinfo {pages}
  {1143} (\bibinfo {year} {2023})}\BibitemShut {NoStop}%
\bibitem [{\citenamefont {Iqbal}\ \emph {et~al.}(2024)\citenamefont {Iqbal},
  \citenamefont {Tantivasadakarn}, \citenamefont {Gatterman}, \citenamefont
  {Gerber}, \citenamefont {Gilmore}, \citenamefont {Gresh}, \citenamefont
  {Hankin}, \citenamefont {Hewitt}, \citenamefont {Horst}, \citenamefont
  {Matheny} \emph {et~al.}}]{Iqbal_2024}%
  \BibitemOpen
  \bibfield  {author} {\bibinfo {author} {\bibfnamefont {M.}~\bibnamefont
  {Iqbal}}, \bibinfo {author} {\bibfnamefont {N.}~\bibnamefont
  {Tantivasadakarn}}, \bibinfo {author} {\bibfnamefont {T.~M.}\ \bibnamefont
  {Gatterman}}, \bibinfo {author} {\bibfnamefont {J.~A.}\ \bibnamefont
  {Gerber}}, \bibinfo {author} {\bibfnamefont {K.}~\bibnamefont {Gilmore}},
  \bibinfo {author} {\bibfnamefont {D.}~\bibnamefont {Gresh}}, \bibinfo
  {author} {\bibfnamefont {A.}~\bibnamefont {Hankin}}, \bibinfo {author}
  {\bibfnamefont {N.}~\bibnamefont {Hewitt}}, \bibinfo {author} {\bibfnamefont
  {C.~V.}\ \bibnamefont {Horst}}, \bibinfo {author} {\bibfnamefont
  {M.}~\bibnamefont {Matheny}}, \emph {et~al.},\ }\bibfield  {title} {\bibinfo
  {title} {Topological order from measurements and feed-forward on a trapped
  ion quantum computer},\ }\href {https://doi.org/10.1038/s42005-024-01698-3}
  {\bibfield  {journal} {\bibinfo  {journal} {Communications Physics}\ }\textbf
  {\bibinfo {volume} {7}},\ \bibinfo {pages} {205} (\bibinfo {year}
  {2024})}\BibitemShut {NoStop}%
\bibitem [{\citenamefont {Bluvstein}\ \emph {et~al.}(2023)\citenamefont
  {Bluvstein}, \citenamefont {Evered}, \citenamefont {Geim}, \citenamefont
  {Li}, \citenamefont {Zhou}, \citenamefont {Manovitz}, \citenamefont {Ebadi},
  \citenamefont {Cain}, \citenamefont {Kalinowski}, \citenamefont {Hangleiter},
  \citenamefont {Bonilla~Ataides}, \citenamefont {Maskara}, \citenamefont
  {Cong}, \citenamefont {Gao}, \citenamefont {Sales~Rodriguez}, \citenamefont
  {Karolyshyn}, \citenamefont {Semeghini}, \citenamefont {Gullans},
  \citenamefont {Greiner}, \citenamefont {Vuletić},\ and\ \citenamefont
  {Lukin}}]{Bluvstein_2023}%
  \BibitemOpen
  \bibfield  {author} {\bibinfo {author} {\bibfnamefont {D.}~\bibnamefont
  {Bluvstein}}, \bibinfo {author} {\bibfnamefont {S.~J.}\ \bibnamefont
  {Evered}}, \bibinfo {author} {\bibfnamefont {A.~A.}\ \bibnamefont {Geim}},
  \bibinfo {author} {\bibfnamefont {S.~H.}\ \bibnamefont {Li}}, \bibinfo
  {author} {\bibfnamefont {H.}~\bibnamefont {Zhou}}, \bibinfo {author}
  {\bibfnamefont {T.}~\bibnamefont {Manovitz}}, \bibinfo {author}
  {\bibfnamefont {S.}~\bibnamefont {Ebadi}}, \bibinfo {author} {\bibfnamefont
  {M.}~\bibnamefont {Cain}}, \bibinfo {author} {\bibfnamefont {M.}~\bibnamefont
  {Kalinowski}}, \bibinfo {author} {\bibfnamefont {D.}~\bibnamefont
  {Hangleiter}}, \bibinfo {author} {\bibfnamefont {J.~P.}\ \bibnamefont
  {Bonilla~Ataides}}, \bibinfo {author} {\bibfnamefont {N.}~\bibnamefont
  {Maskara}}, \bibinfo {author} {\bibfnamefont {I.}~\bibnamefont {Cong}},
  \bibinfo {author} {\bibfnamefont {X.}~\bibnamefont {Gao}}, \bibinfo {author}
  {\bibfnamefont {P.}~\bibnamefont {Sales~Rodriguez}}, \bibinfo {author}
  {\bibfnamefont {T.}~\bibnamefont {Karolyshyn}}, \bibinfo {author}
  {\bibfnamefont {G.}~\bibnamefont {Semeghini}}, \bibinfo {author}
  {\bibfnamefont {M.~J.}\ \bibnamefont {Gullans}}, \bibinfo {author}
  {\bibfnamefont {M.}~\bibnamefont {Greiner}}, \bibinfo {author} {\bibfnamefont
  {V.}~\bibnamefont {Vuletić}},\ and\ \bibinfo {author} {\bibfnamefont
  {M.~D.}\ \bibnamefont {Lukin}},\ }\bibfield  {title} {\bibinfo {title}
  {Logical quantum processor based on reconfigurable atom arrays},\ }\href
  {https://doi.org/10.1038/s41586-023-06927-3} {\bibfield  {journal} {\bibinfo
  {journal} {Nature}\ }\textbf {\bibinfo {volume} {626}},\ \bibinfo {pages}
  {58–65} (\bibinfo {year} {2023})}\BibitemShut {NoStop}%
\bibitem [{\citenamefont {Cai}\ \emph {et~al.}(2023)\citenamefont {Cai},
  \citenamefont {Babbush}, \citenamefont {Benjamin}, \citenamefont {Endo},
  \citenamefont {Huggins}, \citenamefont {Li}, \citenamefont {McClean},\ and\
  \citenamefont {O'Brien}}]{RevModPhys.95.045005}%
  \BibitemOpen
  \bibfield  {author} {\bibinfo {author} {\bibfnamefont {Z.}~\bibnamefont
  {Cai}}, \bibinfo {author} {\bibfnamefont {R.}~\bibnamefont {Babbush}},
  \bibinfo {author} {\bibfnamefont {S.~C.}\ \bibnamefont {Benjamin}}, \bibinfo
  {author} {\bibfnamefont {S.}~\bibnamefont {Endo}}, \bibinfo {author}
  {\bibfnamefont {W.~J.}\ \bibnamefont {Huggins}}, \bibinfo {author}
  {\bibfnamefont {Y.}~\bibnamefont {Li}}, \bibinfo {author} {\bibfnamefont
  {J.~R.}\ \bibnamefont {McClean}},\ and\ \bibinfo {author} {\bibfnamefont
  {T.~E.}\ \bibnamefont {O'Brien}},\ }\bibfield  {title} {\bibinfo {title}
  {Quantum error mitigation},\ }\href
  {https://doi.org/10.1103/RevModPhys.95.045005} {\bibfield  {journal}
  {\bibinfo  {journal} {Rev. Mod. Phys.}\ }\textbf {\bibinfo {volume} {95}},\
  \bibinfo {pages} {045005} (\bibinfo {year} {2023})}\BibitemShut {NoStop}%
\bibitem [{\citenamefont {Kraus}\ and\ \citenamefont
  {Cirac}(2001)}]{PhysRevA.63.062309}%
  \BibitemOpen
  \bibfield  {author} {\bibinfo {author} {\bibfnamefont {B.}~\bibnamefont
  {Kraus}}\ and\ \bibinfo {author} {\bibfnamefont {J.~I.}\ \bibnamefont
  {Cirac}},\ }\bibfield  {title} {\bibinfo {title} {Optimal creation of
  entanglement using a two-qubit gate},\ }\href
  {https://doi.org/10.1103/PhysRevA.63.062309} {\bibfield  {journal} {\bibinfo
  {journal} {Phys. Rev. A}\ }\textbf {\bibinfo {volume} {63}},\ \bibinfo
  {pages} {062309} (\bibinfo {year} {2001})}\BibitemShut {NoStop}%
\bibitem [{\citenamefont {Watts}\ \emph {et~al.}(2015)\citenamefont {Watts},
  \citenamefont {Vala}, \citenamefont {M\"uller}, \citenamefont {Calarco},
  \citenamefont {Whaley}, \citenamefont {Reich}, \citenamefont {Goerz},\ and\
  \citenamefont {Koch}}]{PhysRevA.91.062306}%
  \BibitemOpen
  \bibfield  {author} {\bibinfo {author} {\bibfnamefont {P.}~\bibnamefont
  {Watts}}, \bibinfo {author} {\bibfnamefont {J.~c.~v.}\ \bibnamefont {Vala}},
  \bibinfo {author} {\bibfnamefont {M.~M.}\ \bibnamefont {M\"uller}}, \bibinfo
  {author} {\bibfnamefont {T.}~\bibnamefont {Calarco}}, \bibinfo {author}
  {\bibfnamefont {K.~B.}\ \bibnamefont {Whaley}}, \bibinfo {author}
  {\bibfnamefont {D.~M.}\ \bibnamefont {Reich}}, \bibinfo {author}
  {\bibfnamefont {M.~H.}\ \bibnamefont {Goerz}},\ and\ \bibinfo {author}
  {\bibfnamefont {C.~P.}\ \bibnamefont {Koch}},\ }\bibfield  {title} {\bibinfo
  {title} {Optimizing for an arbitrary perfect entangler. i. functionals},\
  }\href {https://doi.org/10.1103/PhysRevA.91.062306} {\bibfield  {journal}
  {\bibinfo  {journal} {Phys. Rev. A}\ }\textbf {\bibinfo {volume} {91}},\
  \bibinfo {pages} {062306} (\bibinfo {year} {2015})}\BibitemShut {NoStop}%
\bibitem [{\citenamefont {Goerz}\ \emph {et~al.}(2015)\citenamefont {Goerz},
  \citenamefont {Gualdi}, \citenamefont {Reich}, \citenamefont {Koch},
  \citenamefont {Motzoi}, \citenamefont {Whaley}, \citenamefont {Vala},
  \citenamefont {M\"uller}, \citenamefont {Montangero},\ and\ \citenamefont
  {Calarco}}]{PhysRevA.91.062307}%
  \BibitemOpen
  \bibfield  {author} {\bibinfo {author} {\bibfnamefont {M.~H.}\ \bibnamefont
  {Goerz}}, \bibinfo {author} {\bibfnamefont {G.}~\bibnamefont {Gualdi}},
  \bibinfo {author} {\bibfnamefont {D.~M.}\ \bibnamefont {Reich}}, \bibinfo
  {author} {\bibfnamefont {C.~P.}\ \bibnamefont {Koch}}, \bibinfo {author}
  {\bibfnamefont {F.}~\bibnamefont {Motzoi}}, \bibinfo {author} {\bibfnamefont
  {K.~B.}\ \bibnamefont {Whaley}}, \bibinfo {author} {\bibfnamefont {J.~c.~v.}\
  \bibnamefont {Vala}}, \bibinfo {author} {\bibfnamefont {M.~M.}\ \bibnamefont
  {M\"uller}}, \bibinfo {author} {\bibfnamefont {S.}~\bibnamefont
  {Montangero}},\ and\ \bibinfo {author} {\bibfnamefont {T.}~\bibnamefont
  {Calarco}},\ }\bibfield  {title} {\bibinfo {title} {Optimizing for an
  arbitrary perfect entangler. ii. application},\ }\href
  {https://doi.org/10.1103/PhysRevA.91.062307} {\bibfield  {journal} {\bibinfo
  {journal} {Phys. Rev. A}\ }\textbf {\bibinfo {volume} {91}},\ \bibinfo
  {pages} {062307} (\bibinfo {year} {2015})}\BibitemShut {NoStop}%
\bibitem [{\citenamefont {Dehaghani}\ \emph {et~al.}(2025)\citenamefont
  {Dehaghani}, \citenamefont {Aguiar},\ and\ \citenamefont
  {Wisniewski}}]{dehaghani2025optimizingmaximallyentangledstate}%
  \BibitemOpen
  \bibfield  {author} {\bibinfo {author} {\bibfnamefont {N.~B.}\ \bibnamefont
  {Dehaghani}}, \bibinfo {author} {\bibfnamefont {A.~P.}\ \bibnamefont
  {Aguiar}},\ and\ \bibinfo {author} {\bibfnamefont {R.}~\bibnamefont
  {Wisniewski}},\ }\bibfield  {title} {\bibinfo {title} {Optimizing maximally
  entangled state generation via pontryagin's principle},\ }in\ \href
  {https://ieeexplore.ieee.org/document/11158311} {\emph {\bibinfo {booktitle}
  {2025 IEEE International Conference on Quantum Control, Computing and
  Learning (qCCL)}}}\ (\bibinfo {organization} {IEEE},\ \bibinfo {year}
  {2025})\ pp.\ \bibinfo {pages} {112--117}\BibitemShut {NoStop}%
\bibitem [{\citenamefont {Hauschild}\ \emph {et~al.}(2018)\citenamefont
  {Hauschild}, \citenamefont {Leviatan}, \citenamefont {Bardarson},
  \citenamefont {Altman}, \citenamefont {Zaletel},\ and\ \citenamefont
  {Pollmann}}]{PhysRevB.98.235163}%
  \BibitemOpen
  \bibfield  {author} {\bibinfo {author} {\bibfnamefont {J.}~\bibnamefont
  {Hauschild}}, \bibinfo {author} {\bibfnamefont {E.}~\bibnamefont {Leviatan}},
  \bibinfo {author} {\bibfnamefont {J.~H.}\ \bibnamefont {Bardarson}}, \bibinfo
  {author} {\bibfnamefont {E.}~\bibnamefont {Altman}}, \bibinfo {author}
  {\bibfnamefont {M.~P.}\ \bibnamefont {Zaletel}},\ and\ \bibinfo {author}
  {\bibfnamefont {F.}~\bibnamefont {Pollmann}},\ }\bibfield  {title} {\bibinfo
  {title} {Finding purifications with minimal entanglement},\ }\href
  {https://doi.org/10.1103/PhysRevB.98.235163} {\bibfield  {journal} {\bibinfo
  {journal} {Phys. Rev. B}\ }\textbf {\bibinfo {volume} {98}},\ \bibinfo
  {pages} {235163} (\bibinfo {year} {2018})}\BibitemShut {NoStop}%
\bibitem [{\citenamefont {Ljubotina}\ \emph {et~al.}(2022)\citenamefont
  {Ljubotina}, \citenamefont {Roos}, \citenamefont {Abanin},\ and\
  \citenamefont {Serbyn}}]{PRXQuantum.3.030343}%
  \BibitemOpen
  \bibfield  {author} {\bibinfo {author} {\bibfnamefont {M.}~\bibnamefont
  {Ljubotina}}, \bibinfo {author} {\bibfnamefont {B.}~\bibnamefont {Roos}},
  \bibinfo {author} {\bibfnamefont {D.~A.}\ \bibnamefont {Abanin}},\ and\
  \bibinfo {author} {\bibfnamefont {M.}~\bibnamefont {Serbyn}},\ }\bibfield
  {title} {\bibinfo {title} {Optimal steering of matrix product states and
  quantum many-body scars},\ }\href
  {https://doi.org/10.1103/PRXQuantum.3.030343} {\bibfield  {journal} {\bibinfo
   {journal} {PRX Quantum}\ }\textbf {\bibinfo {volume} {3}},\ \bibinfo {pages}
  {030343} (\bibinfo {year} {2022})}\BibitemShut {NoStop}%
\bibitem [{\citenamefont {Lu}\ \emph {et~al.}(2024)\citenamefont {Lu},
  \citenamefont {Shi}, \citenamefont {Wang}, \citenamefont {Hu},\ and\
  \citenamefont {Ran}}]{Lu_2024}%
  \BibitemOpen
  \bibfield  {author} {\bibinfo {author} {\bibfnamefont {Y.}~\bibnamefont
  {Lu}}, \bibinfo {author} {\bibfnamefont {P.}~\bibnamefont {Shi}}, \bibinfo
  {author} {\bibfnamefont {X.-H.}\ \bibnamefont {Wang}}, \bibinfo {author}
  {\bibfnamefont {J.}~\bibnamefont {Hu}},\ and\ \bibinfo {author}
  {\bibfnamefont {S.-J.}\ \bibnamefont {Ran}},\ }\bibfield  {title} {\bibinfo
  {title} {Persistent ballistic entanglement spreading with optimal control in
  quantum spin chains},\ }\href
  {https://doi.org/10.1103/physrevlett.133.070402} {\bibfield  {journal}
  {\bibinfo  {journal} {Physical Review Letters}\ }\textbf {\bibinfo {volume}
  {133}},\ \bibinfo {pages} {070402} (\bibinfo {year} {2024})}\BibitemShut
  {NoStop}%
\bibitem [{\citenamefont {Cramer}\ \emph {et~al.}(2010)\citenamefont {Cramer},
  \citenamefont {Plenio}, \citenamefont {Flammia}, \citenamefont {Somma},
  \citenamefont {Gross}, \citenamefont {Bartlett}, \citenamefont
  {Landon-Cardinal}, \citenamefont {Poulin},\ and\ \citenamefont
  {Liu}}]{Cramer_2010}%
  \BibitemOpen
  \bibfield  {author} {\bibinfo {author} {\bibfnamefont {M.}~\bibnamefont
  {Cramer}}, \bibinfo {author} {\bibfnamefont {M.~B.}\ \bibnamefont {Plenio}},
  \bibinfo {author} {\bibfnamefont {S.~T.}\ \bibnamefont {Flammia}}, \bibinfo
  {author} {\bibfnamefont {R.}~\bibnamefont {Somma}}, \bibinfo {author}
  {\bibfnamefont {D.}~\bibnamefont {Gross}}, \bibinfo {author} {\bibfnamefont
  {S.~D.}\ \bibnamefont {Bartlett}}, \bibinfo {author} {\bibfnamefont
  {O.}~\bibnamefont {Landon-Cardinal}}, \bibinfo {author} {\bibfnamefont
  {D.}~\bibnamefont {Poulin}},\ and\ \bibinfo {author} {\bibfnamefont {Y.-K.}\
  \bibnamefont {Liu}},\ }\bibfield  {title} {\bibinfo {title} {Efficient
  quantum state tomography},\ }\href
  {https://www.nature.com/articles/ncomms1147} {\bibfield  {journal} {\bibinfo
  {journal} {Nature communications}\ }\textbf {\bibinfo {volume} {1}},\
  \bibinfo {pages} {149} (\bibinfo {year} {2010})}\BibitemShut {NoStop}%
\bibitem [{\citenamefont {Tashev}\ \emph {et~al.}(2024)\citenamefont {Tashev},
  \citenamefont {Petrov}, \citenamefont {Metz},\ and\ \citenamefont
  {Bukov}}]{tashev2024reinforcementlearningdisentanglemultiqubit}%
  \BibitemOpen
  \bibfield  {author} {\bibinfo {author} {\bibfnamefont {P.}~\bibnamefont
  {Tashev}}, \bibinfo {author} {\bibfnamefont {S.}~\bibnamefont {Petrov}},
  \bibinfo {author} {\bibfnamefont {F.}~\bibnamefont {Metz}},\ and\ \bibinfo
  {author} {\bibfnamefont {M.}~\bibnamefont {Bukov}},\ }\href
  {https://arxiv.org/abs/2406.07884} {\bibinfo {title} {Reinforcement learning
  to disentangle multiqubit quantum states from partial observations}}
  (\bibinfo {year} {2024}),\ \Eprint {https://arxiv.org/abs/2406.07884}
  {arXiv:2406.07884 [quant-ph]} \BibitemShut {NoStop}%
\bibitem [{\citenamefont {F\"osel}\ \emph {et~al.}(2018)\citenamefont
  {F\"osel}, \citenamefont {Tighineanu}, \citenamefont {Weiss},\ and\
  \citenamefont {Marquardt}}]{PhysRevX.8.031084}%
  \BibitemOpen
  \bibfield  {author} {\bibinfo {author} {\bibfnamefont {T.}~\bibnamefont
  {F\"osel}}, \bibinfo {author} {\bibfnamefont {P.}~\bibnamefont {Tighineanu}},
  \bibinfo {author} {\bibfnamefont {T.}~\bibnamefont {Weiss}},\ and\ \bibinfo
  {author} {\bibfnamefont {F.}~\bibnamefont {Marquardt}},\ }\bibfield  {title}
  {\bibinfo {title} {Reinforcement learning with neural networks for quantum
  feedback},\ }\href {https://doi.org/10.1103/PhysRevX.8.031084} {\bibfield
  {journal} {\bibinfo  {journal} {Phys. Rev. X}\ }\textbf {\bibinfo {volume}
  {8}},\ \bibinfo {pages} {031084} (\bibinfo {year} {2018})}\BibitemShut
  {NoStop}%
\bibitem [{\citenamefont {Bukov}\ \emph {et~al.}(2018)\citenamefont {Bukov},
  \citenamefont {Day}, \citenamefont {Sels}, \citenamefont {Weinberg},
  \citenamefont {Polkovnikov},\ and\ \citenamefont
  {Mehta}}]{PhysRevX.8.031086}%
  \BibitemOpen
  \bibfield  {author} {\bibinfo {author} {\bibfnamefont {M.}~\bibnamefont
  {Bukov}}, \bibinfo {author} {\bibfnamefont {A.~G.~R.}\ \bibnamefont {Day}},
  \bibinfo {author} {\bibfnamefont {D.}~\bibnamefont {Sels}}, \bibinfo {author}
  {\bibfnamefont {P.}~\bibnamefont {Weinberg}}, \bibinfo {author}
  {\bibfnamefont {A.}~\bibnamefont {Polkovnikov}},\ and\ \bibinfo {author}
  {\bibfnamefont {P.}~\bibnamefont {Mehta}},\ }\bibfield  {title} {\bibinfo
  {title} {Reinforcement learning in different phases of quantum control},\
  }\href {https://doi.org/10.1103/PhysRevX.8.031086} {\bibfield  {journal}
  {\bibinfo  {journal} {Phys. Rev. X}\ }\textbf {\bibinfo {volume} {8}},\
  \bibinfo {pages} {031086} (\bibinfo {year} {2018})}\BibitemShut {NoStop}%
\bibitem [{\citenamefont {Bukov}(2018)}]{PhysRevB.98.224305}%
  \BibitemOpen
  \bibfield  {author} {\bibinfo {author} {\bibfnamefont {M.}~\bibnamefont
  {Bukov}},\ }\bibfield  {title} {\bibinfo {title} {Reinforcement learning for
  autonomous preparation of floquet-engineered states: Inverting the quantum
  kapitza oscillator},\ }\href {https://doi.org/10.1103/PhysRevB.98.224305}
  {\bibfield  {journal} {\bibinfo  {journal} {Phys. Rev. B}\ }\textbf {\bibinfo
  {volume} {98}},\ \bibinfo {pages} {224305} (\bibinfo {year}
  {2018})}\BibitemShut {NoStop}%
\bibitem [{\citenamefont {Dalgaard}\ \emph {et~al.}(2020)\citenamefont
  {Dalgaard}, \citenamefont {Motzoi}, \citenamefont {S{\o}rensen},\ and\
  \citenamefont {Sherson}}]{Dalgaard_2020}%
  \BibitemOpen
  \bibfield  {author} {\bibinfo {author} {\bibfnamefont {M.}~\bibnamefont
  {Dalgaard}}, \bibinfo {author} {\bibfnamefont {F.}~\bibnamefont {Motzoi}},
  \bibinfo {author} {\bibfnamefont {J.~J.}\ \bibnamefont {S{\o}rensen}},\ and\
  \bibinfo {author} {\bibfnamefont {J.}~\bibnamefont {Sherson}},\ }\bibfield
  {title} {\bibinfo {title} {Global optimization of quantum dynamics with
  alphazero deep exploration},\ }\href
  {https://www.nature.com/articles/s41534-019-0241-0} {\bibfield  {journal}
  {\bibinfo  {journal} {NPJ quantum information}\ }\textbf {\bibinfo {volume}
  {6}},\ \bibinfo {pages} {6} (\bibinfo {year} {2020})}\BibitemShut {NoStop}%
\bibitem [{\citenamefont {Haug}\ \emph {et~al.}(2020)\citenamefont {Haug},
  \citenamefont {Mok}, \citenamefont {You}, \citenamefont {Zhang},
  \citenamefont {Eng~Png},\ and\ \citenamefont {Kwek}}]{Haug_2021}%
  \BibitemOpen
  \bibfield  {author} {\bibinfo {author} {\bibfnamefont {T.}~\bibnamefont
  {Haug}}, \bibinfo {author} {\bibfnamefont {W.-K.}\ \bibnamefont {Mok}},
  \bibinfo {author} {\bibfnamefont {J.-B.}\ \bibnamefont {You}}, \bibinfo
  {author} {\bibfnamefont {W.}~\bibnamefont {Zhang}}, \bibinfo {author}
  {\bibfnamefont {C.}~\bibnamefont {Eng~Png}},\ and\ \bibinfo {author}
  {\bibfnamefont {L.-C.}\ \bibnamefont {Kwek}},\ }\bibfield  {title} {\bibinfo
  {title} {Classifying global state preparation via deep reinforcement
  learning},\ }\href {https://doi.org/10.1088/2632-2153/abc81f} {\bibfield
  {journal} {\bibinfo  {journal} {Machine Learning: Science and Technology}\
  }\textbf {\bibinfo {volume} {2}},\ \bibinfo {pages} {01LT02} (\bibinfo {year}
  {2020})}\BibitemShut {NoStop}%
\bibitem [{\citenamefont {Mackeprang}\ \emph {et~al.}(2020)\citenamefont
  {Mackeprang}, \citenamefont {Dasari},\ and\ \citenamefont
  {Wrachtrup}}]{Mackeprang_2020}%
  \BibitemOpen
  \bibfield  {author} {\bibinfo {author} {\bibfnamefont {J.}~\bibnamefont
  {Mackeprang}}, \bibinfo {author} {\bibfnamefont {D.~B.~R.}\ \bibnamefont
  {Dasari}},\ and\ \bibinfo {author} {\bibfnamefont {J.}~\bibnamefont
  {Wrachtrup}},\ }\bibfield  {title} {\bibinfo {title} {A reinforcement
  learning approach for quantum state engineering},\ }\href
  {https://doi.org/10.1007/s42484-020-00016-8} {\bibfield  {journal} {\bibinfo
  {journal} {Quantum Machine Intelligence}\ }\textbf {\bibinfo {volume} {2}},\
  \bibinfo {pages} {5} (\bibinfo {year} {2020})}\BibitemShut {NoStop}%
\bibitem [{\citenamefont {Yao}\ \emph {et~al.}(2020)\citenamefont {Yao},
  \citenamefont {Bukov},\ and\ \citenamefont
  {Lin}}]{yao2020policygradientbasedquantum}%
  \BibitemOpen
  \bibfield  {author} {\bibinfo {author} {\bibfnamefont {J.}~\bibnamefont
  {Yao}}, \bibinfo {author} {\bibfnamefont {M.}~\bibnamefont {Bukov}},\ and\
  \bibinfo {author} {\bibfnamefont {L.}~\bibnamefont {Lin}},\ }\bibfield
  {title} {\bibinfo {title} {Policy gradient based quantum approximate
  optimization algorithm},\ }in\ \href
  {https://proceedings.mlr.press/v107/yao20a.html} {\emph {\bibinfo {booktitle}
  {Mathematical and scientific machine learning}}}\ (\bibinfo {organization}
  {PMLR},\ \bibinfo {year} {2020})\ pp.\ \bibinfo {pages}
  {605--634}\BibitemShut {NoStop}%
\bibitem [{\citenamefont {Yao}\ \emph {et~al.}(2022)\citenamefont {Yao},
  \citenamefont {Kottering}, \citenamefont {Gundlach}, \citenamefont {Lin},\
  and\ \citenamefont {Bukov}}]{yao2020noiserobustendtoendquantumcontrol}%
  \BibitemOpen
  \bibfield  {author} {\bibinfo {author} {\bibfnamefont {J.}~\bibnamefont
  {Yao}}, \bibinfo {author} {\bibfnamefont {P.}~\bibnamefont {Kottering}},
  \bibinfo {author} {\bibfnamefont {H.}~\bibnamefont {Gundlach}}, \bibinfo
  {author} {\bibfnamefont {L.}~\bibnamefont {Lin}},\ and\ \bibinfo {author}
  {\bibfnamefont {M.}~\bibnamefont {Bukov}},\ }\bibfield  {title} {\bibinfo
  {title} {Noise-robust end-to-end quantum control using deep autoregressive
  policy networks},\ }in\ \href
  {https://proceedings.mlr.press/v145/yao22a.html} {\emph {\bibinfo {booktitle}
  {Mathematical and scientific machine learning}}}\ (\bibinfo {organization}
  {PMLR},\ \bibinfo {year} {2022})\ pp.\ \bibinfo {pages}
  {1044--1081}\BibitemShut {NoStop}%
\bibitem [{\citenamefont {Guo}\ \emph {et~al.}(2021)\citenamefont {Guo},
  \citenamefont {Chen}, \citenamefont {Liu}, \citenamefont {Xue}, \citenamefont
  {Chen}, \citenamefont {Cao}, \citenamefont {Mao}, \citenamefont {Tey},\ and\
  \citenamefont {You}}]{PhysRevLett.126.060401}%
  \BibitemOpen
  \bibfield  {author} {\bibinfo {author} {\bibfnamefont {S.-F.}\ \bibnamefont
  {Guo}}, \bibinfo {author} {\bibfnamefont {F.}~\bibnamefont {Chen}}, \bibinfo
  {author} {\bibfnamefont {Q.}~\bibnamefont {Liu}}, \bibinfo {author}
  {\bibfnamefont {M.}~\bibnamefont {Xue}}, \bibinfo {author} {\bibfnamefont
  {J.-J.}\ \bibnamefont {Chen}}, \bibinfo {author} {\bibfnamefont {J.-H.}\
  \bibnamefont {Cao}}, \bibinfo {author} {\bibfnamefont {T.-W.}\ \bibnamefont
  {Mao}}, \bibinfo {author} {\bibfnamefont {M.~K.}\ \bibnamefont {Tey}},\ and\
  \bibinfo {author} {\bibfnamefont {L.}~\bibnamefont {You}},\ }\bibfield
  {title} {\bibinfo {title} {Faster state preparation across quantum phase
  transition assisted by reinforcement learning},\ }\href
  {https://doi.org/10.1103/PhysRevLett.126.060401} {\bibfield  {journal}
  {\bibinfo  {journal} {Phys. Rev. Lett.}\ }\textbf {\bibinfo {volume} {126}},\
  \bibinfo {pages} {060401} (\bibinfo {year} {2021})}\BibitemShut {NoStop}%
\bibitem [{\citenamefont {Niu}\ \emph {et~al.}(2019)\citenamefont {Niu},
  \citenamefont {Boixo}, \citenamefont {Smelyanskiy},\ and\ \citenamefont
  {Neven}}]{niu2018universalquantumcontroldeep}%
  \BibitemOpen
  \bibfield  {author} {\bibinfo {author} {\bibfnamefont {M.~Y.}\ \bibnamefont
  {Niu}}, \bibinfo {author} {\bibfnamefont {S.}~\bibnamefont {Boixo}}, \bibinfo
  {author} {\bibfnamefont {V.~N.}\ \bibnamefont {Smelyanskiy}},\ and\ \bibinfo
  {author} {\bibfnamefont {H.}~\bibnamefont {Neven}},\ }\bibfield  {title}
  {\bibinfo {title} {Universal quantum control through deep reinforcement
  learning},\ }\href {https://www.nature.com/articles/s41534-019-0141-3}
  {\bibfield  {journal} {\bibinfo  {journal} {npj Quantum Information}\
  }\textbf {\bibinfo {volume} {5}},\ \bibinfo {pages} {33} (\bibinfo {year}
  {2019})}\BibitemShut {NoStop}%
\bibitem [{\citenamefont {Zhang}\ \emph {et~al.}(2020)\citenamefont {Zhang},
  \citenamefont {Zheng}, \citenamefont {Zhang},\ and\ \citenamefont
  {Deng}}]{PhysRevLett.125.170501}%
  \BibitemOpen
  \bibfield  {author} {\bibinfo {author} {\bibfnamefont {Y.-H.}\ \bibnamefont
  {Zhang}}, \bibinfo {author} {\bibfnamefont {P.-L.}\ \bibnamefont {Zheng}},
  \bibinfo {author} {\bibfnamefont {Y.}~\bibnamefont {Zhang}},\ and\ \bibinfo
  {author} {\bibfnamefont {D.-L.}\ \bibnamefont {Deng}},\ }\bibfield  {title}
  {\bibinfo {title} {Topological quantum compiling with reinforcement
  learning},\ }\href {https://doi.org/10.1103/PhysRevLett.125.170501}
  {\bibfield  {journal} {\bibinfo  {journal} {Phys. Rev. Lett.}\ }\textbf
  {\bibinfo {volume} {125}},\ \bibinfo {pages} {170501} (\bibinfo {year}
  {2020})}\BibitemShut {NoStop}%
\bibitem [{\citenamefont {Bolens}\ and\ \citenamefont
  {Heyl}(2021)}]{PhysRevLett.127.110502}%
  \BibitemOpen
  \bibfield  {author} {\bibinfo {author} {\bibfnamefont {A.}~\bibnamefont
  {Bolens}}\ and\ \bibinfo {author} {\bibfnamefont {M.}~\bibnamefont {Heyl}},\
  }\bibfield  {title} {\bibinfo {title} {Reinforcement learning for digital
  quantum simulation},\ }\href {https://doi.org/10.1103/PhysRevLett.127.110502}
  {\bibfield  {journal} {\bibinfo  {journal} {Phys. Rev. Lett.}\ }\textbf
  {\bibinfo {volume} {127}},\ \bibinfo {pages} {110502} (\bibinfo {year}
  {2021})}\BibitemShut {NoStop}%
\bibitem [{\citenamefont {Moro}\ \emph {et~al.}(2021)\citenamefont {Moro},
  \citenamefont {Paris}, \citenamefont {Restelli},\ and\ \citenamefont
  {Prati}}]{Moro_2021}%
  \BibitemOpen
  \bibfield  {author} {\bibinfo {author} {\bibfnamefont {L.}~\bibnamefont
  {Moro}}, \bibinfo {author} {\bibfnamefont {M.~G.}\ \bibnamefont {Paris}},
  \bibinfo {author} {\bibfnamefont {M.}~\bibnamefont {Restelli}},\ and\
  \bibinfo {author} {\bibfnamefont {E.}~\bibnamefont {Prati}},\ }\bibfield
  {title} {\bibinfo {title} {Quantum compiling by deep reinforcement
  learning},\ }\href {https://doi.org/10.1038/s42005-021-00684-3} {\bibfield
  {journal} {\bibinfo  {journal} {Communications Physics}\ }\textbf {\bibinfo
  {volume} {4}},\ \bibinfo {pages} {178} (\bibinfo {year} {2021})}\BibitemShut
  {NoStop}%
\bibitem [{\citenamefont {He}\ \emph {et~al.}(2021)\citenamefont {He},
  \citenamefont {Li}, \citenamefont {Zheng}, \citenamefont {Li},\ and\
  \citenamefont {Situ}}]{He_2021}%
  \BibitemOpen
  \bibfield  {author} {\bibinfo {author} {\bibfnamefont {Z.}~\bibnamefont
  {He}}, \bibinfo {author} {\bibfnamefont {L.}~\bibnamefont {Li}}, \bibinfo
  {author} {\bibfnamefont {S.}~\bibnamefont {Zheng}}, \bibinfo {author}
  {\bibfnamefont {Y.}~\bibnamefont {Li}},\ and\ \bibinfo {author}
  {\bibfnamefont {H.}~\bibnamefont {Situ}},\ }\bibfield  {title} {\bibinfo
  {title} {Variational quantum compiling with double q-learning},\ }\href
  {https://doi.org/10.1088/1367-2630/abe0ae} {\bibfield  {journal} {\bibinfo
  {journal} {New Journal of Physics}\ }\textbf {\bibinfo {volume} {23}},\
  \bibinfo {pages} {033002} (\bibinfo {year} {2021})}\BibitemShut {NoStop}%
\bibitem [{\citenamefont {Fösel}\ \emph {et~al.}(2021)\citenamefont {Fösel},
  \citenamefont {Niu}, \citenamefont {Marquardt},\ and\ \citenamefont
  {Li}}]{fösel2021quantumcircuitoptimizationdeep}%
  \BibitemOpen
  \bibfield  {author} {\bibinfo {author} {\bibfnamefont {T.}~\bibnamefont
  {Fösel}}, \bibinfo {author} {\bibfnamefont {M.~Y.}\ \bibnamefont {Niu}},
  \bibinfo {author} {\bibfnamefont {F.}~\bibnamefont {Marquardt}},\ and\
  \bibinfo {author} {\bibfnamefont {L.}~\bibnamefont {Li}},\ }\href
  {https://arxiv.org/abs/2103.07585} {\bibinfo {title} {Quantum circuit
  optimization with deep reinforcement learning}} (\bibinfo {year} {2021}),\
  \Eprint {https://arxiv.org/abs/2103.07585} {arXiv:2103.07585 [quant-ph]}
  \BibitemShut {NoStop}%
\bibitem [{\citenamefont {Meyer}\ \emph {et~al.}(2024)\citenamefont {Meyer},
  \citenamefont {Ufrecht}, \citenamefont {Periyasamy}, \citenamefont {Scherer},
  \citenamefont {Plinge},\ and\ \citenamefont
  {Mutschler}}]{meyer2024surveyquantumreinforcementlearning}%
  \BibitemOpen
  \bibfield  {author} {\bibinfo {author} {\bibfnamefont {N.}~\bibnamefont
  {Meyer}}, \bibinfo {author} {\bibfnamefont {C.}~\bibnamefont {Ufrecht}},
  \bibinfo {author} {\bibfnamefont {M.}~\bibnamefont {Periyasamy}}, \bibinfo
  {author} {\bibfnamefont {D.~D.}\ \bibnamefont {Scherer}}, \bibinfo {author}
  {\bibfnamefont {A.}~\bibnamefont {Plinge}},\ and\ \bibinfo {author}
  {\bibfnamefont {C.}~\bibnamefont {Mutschler}},\ }\href
  {https://arxiv.org/abs/2211.03464} {\bibinfo {title} {A survey on quantum
  reinforcement learning}} (\bibinfo {year} {2024}),\ \Eprint
  {https://arxiv.org/abs/2211.03464} {arXiv:2211.03464 [quant-ph]} \BibitemShut
  {NoStop}%
\bibitem [{\citenamefont {Dong}\ \emph {et~al.}(2008)\citenamefont {Dong},
  \citenamefont {Chen}, \citenamefont {Li},\ and\ \citenamefont
  {Tarn}}]{4579244}%
  \BibitemOpen
  \bibfield  {author} {\bibinfo {author} {\bibfnamefont {D.}~\bibnamefont
  {Dong}}, \bibinfo {author} {\bibfnamefont {C.}~\bibnamefont {Chen}}, \bibinfo
  {author} {\bibfnamefont {H.}~\bibnamefont {Li}},\ and\ \bibinfo {author}
  {\bibfnamefont {T.-J.}\ \bibnamefont {Tarn}},\ }\bibfield  {title} {\bibinfo
  {title} {Quantum reinforcement learning},\ }\href
  {https://doi.org/10.1109/TSMCB.2008.925743} {\bibfield  {journal} {\bibinfo
  {journal} {IEEE Transactions on Systems, Man, and Cybernetics, Part B
  (Cybernetics)}\ }\textbf {\bibinfo {volume} {38}},\ \bibinfo {pages} {1207}
  (\bibinfo {year} {2008})}\BibitemShut {NoStop}%
\bibitem [{\citenamefont {Alexeev}\ \emph {et~al.}(2025)\citenamefont
  {Alexeev}, \citenamefont {Farag}, \citenamefont {Patti}, \citenamefont
  {Wolf}, \citenamefont {Ares}, \citenamefont {Aspuru-Guzik}, \citenamefont
  {Benjamin}, \citenamefont {Cai}, \citenamefont {Cao}, \citenamefont
  {Chamberland} \emph {et~al.}}]{alexeev2025artificial}%
  \BibitemOpen
  \bibfield  {author} {\bibinfo {author} {\bibfnamefont {Y.}~\bibnamefont
  {Alexeev}}, \bibinfo {author} {\bibfnamefont {M.~H.}\ \bibnamefont {Farag}},
  \bibinfo {author} {\bibfnamefont {T.~L.}\ \bibnamefont {Patti}}, \bibinfo
  {author} {\bibfnamefont {M.~E.}\ \bibnamefont {Wolf}}, \bibinfo {author}
  {\bibfnamefont {N.}~\bibnamefont {Ares}}, \bibinfo {author} {\bibfnamefont
  {A.}~\bibnamefont {Aspuru-Guzik}}, \bibinfo {author} {\bibfnamefont {S.~C.}\
  \bibnamefont {Benjamin}}, \bibinfo {author} {\bibfnamefont {Z.}~\bibnamefont
  {Cai}}, \bibinfo {author} {\bibfnamefont {S.}~\bibnamefont {Cao}}, \bibinfo
  {author} {\bibfnamefont {C.}~\bibnamefont {Chamberland}}, \emph {et~al.},\
  }\bibfield  {title} {\bibinfo {title} {Artificial intelligence for quantum
  computing},\ }\href {https://www.nature.com/articles/s41467-025-65836-3}
  {\bibfield  {journal} {\bibinfo  {journal} {Nature Communications}\ }\textbf
  {\bibinfo {volume} {16}},\ \bibinfo {pages} {10829} (\bibinfo {year}
  {2025})}\BibitemShut {NoStop}%
\bibitem [{\citenamefont {Jerbi}\ \emph {et~al.}(2021)\citenamefont {Jerbi},
  \citenamefont {Gyurik}, \citenamefont {Marshall}, \citenamefont {Briegel},\
  and\ \citenamefont {Dunjko}}]{NEURIPS2021_eec96a7f}%
  \BibitemOpen
  \bibfield  {author} {\bibinfo {author} {\bibfnamefont {S.}~\bibnamefont
  {Jerbi}}, \bibinfo {author} {\bibfnamefont {C.}~\bibnamefont {Gyurik}},
  \bibinfo {author} {\bibfnamefont {S.}~\bibnamefont {Marshall}}, \bibinfo
  {author} {\bibfnamefont {H.}~\bibnamefont {Briegel}},\ and\ \bibinfo {author}
  {\bibfnamefont {V.}~\bibnamefont {Dunjko}},\ }\bibfield  {title} {\bibinfo
  {title} {Parametrized quantum policies for reinforcement learning},\ }in\
  \href
  {https://proceedings.neurips.cc/paper_files/paper/2021/file/eec96a7f788e88184c0e713456026f3f-Paper.pdf}
  {\emph {\bibinfo {booktitle} {Advances in Neural Information Processing
  Systems}}},\ Vol.~\bibinfo {volume} {34},\ \bibinfo {editor} {edited by\
  \bibinfo {editor} {\bibfnamefont {M.}~\bibnamefont {Ranzato}}, \bibinfo
  {editor} {\bibfnamefont {A.}~\bibnamefont {Beygelzimer}}, \bibinfo {editor}
  {\bibfnamefont {Y.}~\bibnamefont {Dauphin}}, \bibinfo {editor} {\bibfnamefont
  {P.}~\bibnamefont {Liang}},\ and\ \bibinfo {editor} {\bibfnamefont {J.~W.}\
  \bibnamefont {Vaughan}}}\ (\bibinfo  {publisher} {Curran Associates, Inc.},\
  \bibinfo {year} {2021})\ pp.\ \bibinfo {pages} {28362--28375}\BibitemShut
  {NoStop}%
\bibitem [{\citenamefont {Skolik}\ \emph {et~al.}(2022)\citenamefont {Skolik},
  \citenamefont {Jerbi},\ and\ \citenamefont {Dunjko}}]{Skolik_2022}%
  \BibitemOpen
  \bibfield  {author} {\bibinfo {author} {\bibfnamefont {A.}~\bibnamefont
  {Skolik}}, \bibinfo {author} {\bibfnamefont {S.}~\bibnamefont {Jerbi}},\ and\
  \bibinfo {author} {\bibfnamefont {V.}~\bibnamefont {Dunjko}},\ }\bibfield
  {title} {\bibinfo {title} {Quantum agents in the gym: a variational quantum
  algorithm for deep q-learning},\ }\href
  {https://doi.org/10.22331/q-2022-05-24-720} {\bibfield  {journal} {\bibinfo
  {journal} {Quantum}\ }\textbf {\bibinfo {volume} {6}},\ \bibinfo {pages}
  {720} (\bibinfo {year} {2022})}\BibitemShut {NoStop}%
\bibitem [{\citenamefont {Jin}\ \emph {et~al.}(2025)\citenamefont {Jin},
  \citenamefont {Wang}, \citenamefont {Xu}, \citenamefont {Zhuang},
  \citenamefont {Hu},\ and\ \citenamefont
  {Liu}}]{jin2025ppoqproximalpolicyoptimization}%
  \BibitemOpen
  \bibfield  {author} {\bibinfo {author} {\bibfnamefont {Y.-X.}\ \bibnamefont
  {Jin}}, \bibinfo {author} {\bibfnamefont {Z.-W.}\ \bibnamefont {Wang}},
  \bibinfo {author} {\bibfnamefont {H.-Z.}\ \bibnamefont {Xu}}, \bibinfo
  {author} {\bibfnamefont {W.-F.}\ \bibnamefont {Zhuang}}, \bibinfo {author}
  {\bibfnamefont {M.-J.}\ \bibnamefont {Hu}},\ and\ \bibinfo {author}
  {\bibfnamefont {D.~E.}\ \bibnamefont {Liu}},\ }\href
  {https://arxiv.org/abs/2501.07085} {\bibinfo {title} {Ppo-q: Proximal policy
  optimization with parametrized quantum policies or values}} (\bibinfo {year}
  {2025}),\ \Eprint {https://arxiv.org/abs/2501.07085} {arXiv:2501.07085
  [quant-ph]} \BibitemShut {NoStop}%
\bibitem [{\citenamefont {Wu}\ \emph {et~al.}(2025)\citenamefont {Wu},
  \citenamefont {Jin}, \citenamefont {Wen}, \citenamefont {Han},\ and\
  \citenamefont {Wang}}]{Wu_2025}%
  \BibitemOpen
  \bibfield  {author} {\bibinfo {author} {\bibfnamefont {S.}~\bibnamefont
  {Wu}}, \bibinfo {author} {\bibfnamefont {S.}~\bibnamefont {Jin}}, \bibinfo
  {author} {\bibfnamefont {D.}~\bibnamefont {Wen}}, \bibinfo {author}
  {\bibfnamefont {D.}~\bibnamefont {Han}},\ and\ \bibinfo {author}
  {\bibfnamefont {X.}~\bibnamefont {Wang}},\ }\bibfield  {title} {\bibinfo
  {title} {Quantum reinforcement learning in continuous action space},\ }\href
  {https://doi.org/10.22331/q-2025-03-12-1660} {\bibfield  {journal} {\bibinfo
  {journal} {Quantum}\ }\textbf {\bibinfo {volume} {9}},\ \bibinfo {pages}
  {1660} (\bibinfo {year} {2025})}\BibitemShut {NoStop}%
\bibitem [{\citenamefont {Li}\ \emph {et~al.}(2020)\citenamefont {Li},
  \citenamefont {Wallace}, \citenamefont {Shen}, \citenamefont {Lin},
  \citenamefont {Keutzer}, \citenamefont {Klein},\ and\ \citenamefont
  {Gonzalez}}]{li2020trainlargecompressrethinking}%
  \BibitemOpen
  \bibfield  {author} {\bibinfo {author} {\bibfnamefont {Z.}~\bibnamefont
  {Li}}, \bibinfo {author} {\bibfnamefont {E.}~\bibnamefont {Wallace}},
  \bibinfo {author} {\bibfnamefont {S.}~\bibnamefont {Shen}}, \bibinfo {author}
  {\bibfnamefont {K.}~\bibnamefont {Lin}}, \bibinfo {author} {\bibfnamefont
  {K.}~\bibnamefont {Keutzer}}, \bibinfo {author} {\bibfnamefont
  {D.}~\bibnamefont {Klein}},\ and\ \bibinfo {author} {\bibfnamefont
  {J.}~\bibnamefont {Gonzalez}},\ }\bibfield  {title} {\bibinfo {title} {Train
  big, then compress: Rethinking model size for efficient training and
  inference of transformers},\ }in\ \href
  {https://proceedings.mlr.press/v119/li20m.html} {\emph {\bibinfo {booktitle}
  {International Conference on machine learning}}}\ (\bibinfo {organization}
  {PMLR},\ \bibinfo {year} {2020})\ pp.\ \bibinfo {pages}
  {5958--5968}\BibitemShut {NoStop}%
\bibitem [{\citenamefont {Lockwood}\ and\ \citenamefont
  {Si}(2020)}]{10.5555/3505464.3505499}%
  \BibitemOpen
  \bibfield  {author} {\bibinfo {author} {\bibfnamefont {O.}~\bibnamefont
  {Lockwood}}\ and\ \bibinfo {author} {\bibfnamefont {M.}~\bibnamefont {Si}},\
  }\bibfield  {title} {\bibinfo {title} {Reinforcement learning with quantum
  variational circuit},\ }in\ \href
  {https://aaai.org/papers/00245-7437-reinforcement-learning-with-quantum-variational-circuit/}
  {\emph {\bibinfo {booktitle} {Proceedings of the AAAI conference on
  artificial intelligence and interactive digital entertainment}}},\
  Vol.~\bibinfo {volume} {16}\ (\bibinfo {year} {2020})\ pp.\ \bibinfo {pages}
  {245--251}\BibitemShut {NoStop}%
\bibitem [{\citenamefont {Chen}\ \emph {et~al.}(2020)\citenamefont {Chen},
  \citenamefont {Yang}, \citenamefont {Qi}, \citenamefont {Chen}, \citenamefont
  {Ma},\ and\ \citenamefont {Goan}}]{9144562}%
  \BibitemOpen
  \bibfield  {author} {\bibinfo {author} {\bibfnamefont {S.~Y.-C.}\
  \bibnamefont {Chen}}, \bibinfo {author} {\bibfnamefont {C.-H.~H.}\
  \bibnamefont {Yang}}, \bibinfo {author} {\bibfnamefont {J.}~\bibnamefont
  {Qi}}, \bibinfo {author} {\bibfnamefont {P.-Y.}\ \bibnamefont {Chen}},
  \bibinfo {author} {\bibfnamefont {X.}~\bibnamefont {Ma}},\ and\ \bibinfo
  {author} {\bibfnamefont {H.-S.}\ \bibnamefont {Goan}},\ }\bibfield  {title}
  {\bibinfo {title} {Variational quantum circuits for deep reinforcement
  learning},\ }\href {https://doi.org/10.1109/ACCESS.2020.3010470} {\bibfield
  {journal} {\bibinfo  {journal} {IEEE Access}\ }\textbf {\bibinfo {volume}
  {8}},\ \bibinfo {pages} {141007} (\bibinfo {year} {2020})}\BibitemShut
  {NoStop}%
\bibitem [{\citenamefont {Terno}(1999)}]{PhysRevA.59.3320}%
  \BibitemOpen
  \bibfield  {author} {\bibinfo {author} {\bibfnamefont {D.~R.}\ \bibnamefont
  {Terno}},\ }\bibfield  {title} {\bibinfo {title} {Nonlinear operations in
  quantum-information theory},\ }\href
  {https://doi.org/10.1103/PhysRevA.59.3320} {\bibfield  {journal} {\bibinfo
  {journal} {Phys. Rev. A}\ }\textbf {\bibinfo {volume} {59}},\ \bibinfo
  {pages} {3320} (\bibinfo {year} {1999})}\BibitemShut {NoStop}%
\bibitem [{\citenamefont {Kaelbling}\ \emph {et~al.}(1998)\citenamefont
  {Kaelbling}, \citenamefont {Littman},\ and\ \citenamefont
  {Cassandra}}]{KAELBLING199899}%
  \BibitemOpen
  \bibfield  {author} {\bibinfo {author} {\bibfnamefont {L.~P.}\ \bibnamefont
  {Kaelbling}}, \bibinfo {author} {\bibfnamefont {M.~L.}\ \bibnamefont
  {Littman}},\ and\ \bibinfo {author} {\bibfnamefont {A.~R.}\ \bibnamefont
  {Cassandra}},\ }\bibfield  {title} {\bibinfo {title} {Planning and acting in
  partially observable stochastic domains},\ }\href
  {https://doi.org/https://doi.org/10.1016/S0004-3702(98)00023-X} {\bibfield
  {journal} {\bibinfo  {journal} {Artificial Intelligence}\ }\textbf {\bibinfo
  {volume} {101}},\ \bibinfo {pages} {99} (\bibinfo {year} {1998})}\BibitemShut
  {NoStop}%
\bibitem [{\citenamefont {Dong}\ and\ \citenamefont
  {Petersen}(2022)}]{DONG2022243}%
  \BibitemOpen
  \bibfield  {author} {\bibinfo {author} {\bibfnamefont {D.}~\bibnamefont
  {Dong}}\ and\ \bibinfo {author} {\bibfnamefont {I.~R.}\ \bibnamefont
  {Petersen}},\ }\bibfield  {title} {\bibinfo {title} {Quantum estimation,
  control and learning: Opportunities and challenges},\ }\href
  {https://doi.org/https://doi.org/10.1016/j.arcontrol.2022.04.011} {\bibfield
  {journal} {\bibinfo  {journal} {Annual Reviews in Control}\ }\textbf
  {\bibinfo {volume} {54}},\ \bibinfo {pages} {243} (\bibinfo {year}
  {2022})}\BibitemShut {NoStop}%
\bibitem [{\citenamefont {Sotnikov}\ \emph {et~al.}(2022)\citenamefont
  {Sotnikov}, \citenamefont {Iakovlev}, \citenamefont {Iliasov}, \citenamefont
  {Katsnelson}, \citenamefont {Bagrov},\ and\ \citenamefont
  {Mazurenko}}]{Sotnikov_2022}%
  \BibitemOpen
  \bibfield  {author} {\bibinfo {author} {\bibfnamefont {O.}~\bibnamefont
  {Sotnikov}}, \bibinfo {author} {\bibfnamefont {I.}~\bibnamefont {Iakovlev}},
  \bibinfo {author} {\bibfnamefont {A.}~\bibnamefont {Iliasov}}, \bibinfo
  {author} {\bibfnamefont {M.}~\bibnamefont {Katsnelson}}, \bibinfo {author}
  {\bibfnamefont {A.}~\bibnamefont {Bagrov}},\ and\ \bibinfo {author}
  {\bibfnamefont {V.}~\bibnamefont {Mazurenko}},\ }\bibfield  {title} {\bibinfo
  {title} {Certification of quantum states with hidden structure of their
  bitstrings},\ }\href {https://doi.org/10.1038/s41534-022-00559-7} {\bibfield
  {journal} {\bibinfo  {journal} {npj Quantum Information}\ }\textbf {\bibinfo
  {volume} {8}},\ \bibinfo {pages} {41} (\bibinfo {year} {2022})}\BibitemShut
  {NoStop}%
\end{thebibliography}%

\end{document}